\documentclass[preprint,superscriptaddress,eqsecnum,nofootinbib]{revtex4}
\usepackage{graphicx,amssymb,amsbsy,bm,amsmath,psfrag,dsfont}
\usepackage{hyperref}
\usepackage[dvips]{color}

\begin{document}

\newcommand{\be}{\begin{equation}}
\newcommand{\ee}{\end{equation}}
\newcommand{\ba}{\begin{array}}
\newcommand{\ea}{\end{array}}
\newcommand{\beq}[1]{\begin{equation}\label{#1}}
\newcommand{\eeq}{\end{equation}}
\newcommand{\bea}{\begin{eqnarray}}
\newcommand{\eea}{\end{eqnarray}}
\newcommand{\no}{\noindent}
\newcommand{\rf}[1]{(\ref{#1})}
\newcommand{\rarr}{\rightarrow}

\newcommand{\la}{\lambda}
\newcommand{\si}{\sigma}
\newcommand{\vk}{\vec{k}}
\newcommand{\vx}{\vec{x}}
\newcommand{\rvec}{\vec{r}}
\newcommand{\om}{\omega}
\newcommand{\Om}{\Omega}
\newcommand{\ga}{\gamma}
\newcommand{\Ga}{\Gamma}
\newcommand{\gaa}{\Gamma_a}
\newcommand{\al}{\alpha}
\newcommand{\bt}{\beta}
\newcommand{\ep}{\epsilon}
\newcommand{\app}{\approx}
\newcommand{\uvk}{\widehat{\bf{k}}}
\newcommand{\OM}{\overline{M}}
\newcommand{\Det}{\rm{Det}}
\newcommand{\ustar}{u_\star}
\newcommand{\taustar}{\tau_\star}
\newcommand{\PD}{ P_{\rm D} }
\newcommand{\PP}{ P_{\rm P} }
\newcommand{\duzero}{\dot{u}_0}
\newcommand{\dxzero}{\dot{x}_0}
\newcommand{\bu}{\duzero^{\max}}
\newcommand{\nd}{n_\delta}
\newcommand{\nbar}{\bar{n}}
\newcommand{\Fb}[1]{|F(#1)|^2}
\newcommand{\thet}{\theta_{h\bar{\phi}_{\nbar}}}
\newcommand{\half}{\frac{1}{2}}
\newcommand{\quarter}{\frac{1}{4}}

\title{Causality-Violating Higgs Singlets at the LHC}
\author{Chiu Man Ho}\email{chiuman.ho@vanderbilt.edu}
  \affiliation{Department of
  Physics and Astronomy, Vanderbilt University, Nashville, TN 37235, USA}
\author{Thomas J. Weiler}\email{tom.weiler@vanderbilt.edu}
\affiliation{Department of
  Physics and Astronomy, Vanderbilt University, Nashville, TN 37235, USA}

\date{\today}

\begin{abstract}
We construct a simple class of compactified five-dimensional metrics which admits closed
timelike curves (CTCs), and derive the resulting CTCs as analytic solutions to the geodesic equations of motion.
The associated Einstein tensor satisfies all the null, weak, strong and dominant energy conditions.
In particular, no negative-energy ``tachyonic'' matter is required.
In extra-dimensional models where gauge charges are bound to our brane, it is the Kaluza-Klein (KK) modes of
gauge-singlets that may travel through the CTCs.
From our brane point of view, many of these KK modes would appear to travel backward in time.
We give a simple model in which time-traveling Higgs singlets can be
produced by the LHC,
either from decay of the Standard Model (SM) Higgs or through mixing with the SM Higgs.
The signature of these time-traveling singlets is a secondary decay vertex
pre-appearing before the primary vertex which produced them.
The two vertices are correlated by momentum conservation.
We demonstrate that pre-appearing vertices in the Higgs singlet-doublet mixing model may well be
observable at the LHC.

%
%
\end{abstract}

\maketitle

\section{Introduction}

Time travel has always been an ambitious dream in science fiction. However, the possibility of
building a time machine could not even be formulated as science until the discovery of special
and general relativity by Einstein. From the early days of general relativity onward,
theoretical physicists have realized that closed timelike curves (CTCs) are allowed solutions of general
relativity, and hence time travel is theoretically possible.
Many proposals for CTCs in a familiar four-dimensional Universe have been discussed in the literature.
In chronological order (perhaps not the best listing scheme for CTC proposals), proposals include
van Stockum's rotating cylinder~\cite{vanStockum} (extended much later by Tipler~\cite{Tipler}),
G\"{o}del's rotating universe \cite{Godel},
Wheeler's spacetime foam \cite{Wheeler},
Kerr and Kerr-Newman's black hole event horizon interior \cite{Kerr},
Morris, Throne and Yurtsever's traversable wormholes \cite{MTY},
Gott's pair of spinning cosmic strings \cite{Gott},
Alcubierre's warp drive \cite{warp},
and Ori's vacuum torus \cite{Ori}.
More additions to the possibilities continue to unfold~\cite{Gron}.

Common pathologies associated with these candidate CTCs are that the required matter distributions are often unphysical,
tachyonic, unstable under the back-reaction of the metric,
or violate one or more of the desirable null, weak, strong and dominant energy conditions \cite{Visser}.
These common pathologies have led Hawking to formulate his
``chronology protection conjecture"~\cite{Hawking},
which states that even for CTCs allowed by general relativity, some fundamental
law of physics forbids their existence so as to maintain the chronological order of physical processes.
The empirical basis for the conjecture is that so far the human species has not observed non-causal processes.
The logical basis for the conjecture is that we do not know how to make sense of a non-causal Universe.

The possibility of time travel leads to many paradoxes.
The most famous paradoxes include the ``Grandfather'' and ``Bootstrap'' paradoxes \cite{Visser}.
In the Grandfather paradox, one can destroy the necessary initial conditions that lead to one's very existence;
while in the Bootstrap paradox, an effect can be its own cause.
A further paradox is the apparent loss of unitarity, as particles may appear ``now'', having disappeared
at another time ``then'', and vice versa.
However, after almost two decades of intensive research on this subject,
Hawking's conjecture remains a hope that is not mathematically compelling.
For example, it has been shown that there are points on the chronology horizon where the semiclassical Einstein
field equations, on which Hawking's conjecture is based, fail to hold \cite{Wald}.
This and related issues have led many physicists to believe that the validity of chronology protection will
not be settled
until we have a much better understanding of gravity itself, whether quantizable or emergent.
In related work, some aspects of chronology protection in string theory have been studied
in~\cite{Horava,Dyson,AdS,Johnson}.

Popular for the previous decade has been the idea of Arkani-Hamed-Dimopoulos-Dvali (ADD)~\cite{ADD}
that the weakness of gravity on our 4D brane might be explained by large extra dimensions.
A lowered Planck mass is accommodated, with field strengths diluted by the extra dimensions as given by Gauss's Law.
The hope is that low-scale gravity may ameliorate or explain the otherwise fine-tuned hierarchy ratio
$M_{\rm weak}/M_{\rm Planck}$.
In the ADD scenario, all particles with gauge charge, which includes all of the standard model (SM) particles,
are open strings with charged endpoints confined to the brane (our 4D spacetime).
Gauge singlets, which include the graviton,
are closed strings which may freely propagate throughout the brane and bulk (the extra dimensions).
After all, wherever there is spacetime, whether brane or bulk, there is Einstein's gravity.
Gauge singlets other than the graviton are speculative.
They may include sterile neutrinos and scalar singlets.
Due to mixing with gauge non-singlet particles, e.g. active neutrinos or SM Higgs doublets, respectively,
sterile neutrinos or scalar singlets will attain a non-gravitational presence when they traverse the brane.

A generic feature in this ADD picture is the possibility of gauge singlets taking
``shortcuts" through the extra dimensions~\cite{shortcut1,shortcut2,shortcut3,shortcut4,shortcut5},
leading to superluminal communications from the brane's point of view.
The extra dimensions could also be warped~\cite{RS}.
Of particular interest are the so-called asymmetrically warped spacetimes~\cite{csaki}
in which space and time coordinates carry different warp factors.
%
Scenarios of large extra dimensions with asymmetrically warped spacetimes are endowed with superluminal
travel --- a signal, say a graviton, from one point on the brane can take a ``shortcut"
through the bulk and later intersect the brane at a different point,
with a shorter transit time than that of a photon traveling between the
same two points along a brane geodesic.
This suggests that regions that are traditionally ``outside the horizon" could be causally related
by gravitons or other gauge singlets.
Exactly this mechanism has been invoked as a solution to the cosmological horizon problem
without inflation~\cite{freese}.
Although this leads to an apparent causality violation from the brane's point of view, the full 5D theory
may be completely causal.
Superluminal travel through extra-dimensional ``shortcuts'' generally doesn't guarantee a CTC.
To obtain a CTC, one needs the light cone in a \,$t$-versus-$r$ diagram to tip below the
horizontal $r$ axis for part of the path.
Then, for this part of the path, travel along $r$ is truly progressing along negative time.
When the positive time part of the path is added, one has a CTC if the net travel time is negative.

Recently, there was an exploratory attempt to find a CTC using a spacetime with two asymmetrically
warped extra dimensions~\cite{Tom}.  In this work, it was demonstrated that
paths exist which in fact are CTCs.
However, these constructed paths are not solutions of geodesic equations.
The construction demonstrated the existence of CTCs in principle for a class of extra dimensional metrics,
but did not present CTCs which would actually be traversed by particles.
Since geodesic paths minimize the action obtained from a metric,
the conjecture in~\cite{Tom} was that the same action that admits constructed paths with negative or zero time,
admits geodesic paths with even greater negative (or zero) time.

The ambitions of this article are threefold:\,
First of all, we seek a class of CTCs embedded in a single compactified extra dimension.
We require the CTCs to be geodesic paths, so that physical particles will become negative-time travelers.
Secondly, we ensure that this class of CTCs is free of undesirable pathologies.
Thirdly, we ask whether particles traversing these CTC geodesics may reveal unique signatures
in large detectors such as ATLAS and CMS at the LHC.

As we demonstrate in this article, we have successfully found a class of 5D metrics which generates exactly solvable
geodesic equations whose solutions are in fact CTCs.  We adopt an ADD framework where only gauge singlet particles
(gravitons, sterile neutrinos, and Higgs singlets) may leave our 4D brane and traverse the
CTC embedded in the extra dimension.
In this way, the standard paradoxes (described below) are ameliorated,
as no macro objects can get transported back in time.
Scalar gauge-singlets, e.g. Higgs singlets,
mixed or unmixed with their gauge non-singlet siblings, e.g. SM Higgs doublets,
may be produced and detected at the LHC.
The signature of negative-time travel is the appearance of a secondary decay or scattering vertex
{\sl earlier in time} than the occurrence of the primary vertex which produces the time-traveling particle.
The two vertices are associated by overall momentum conservation.

Realizing that the Grandfather, Bootstrap, and Unitarity paradoxes may be logically disturbing,
we now discuss the paradoxes briefly.
First of all, it bears repeating that in the ADD picture, it is only gauge-singlet particles that may travel CTCs.
No claims of human or robot transport backwards through time are made.
And while the paradoxes are unsettling, as was/is quantum mechanics,
we think that it is naive to preclude the possibility of time travel on the grounds of human argument/preference.
The paradoxes may be but seeming contradictions resulting from our ignorance of some fundamental laws of physics
which in fact enforce  consistency~\cite{Novikov:1989sd}.
For instance, in Feynman's path integral language, one should sum over all possible globally defined histories.
It is possible that histories leading to paradoxes may contribute little or nothing to this sum.
In other words, while the Grandfather paradox is dynamically allowed by Einstein's field equations,
it may be kinematically forbidden due to the inaccessibility of self-contradicting histories
in the path integral~\cite{Carlini,Echeverria:1991nk,Friedman:1990xc,Friedman:1992jc,Boulware:1992pm,Mironov:2007bm}.
In the Bootstrap paradox, the information, events, or objects in the causal loop do
not seem to have an identifiable cause.
The entities appear as if they were eternally existing, with the causation being pushed back to the infinite past.
But the logic of the Bootstrap paradox does not seem to preclude the possibility of time travel in any compelling manner.

The Unitarity paradox is unsettling as
it seems to suggest that the past can get particles from the future ``for free''.
If Nature respects unitarity as one of her most fundamental principles, she may have a consistent
way (unknown at present) to implement it even in the face of causality violation.
It is also conceivable that Nature sacrifices unitarity.
Precedent seems to exist in quantum mechanics:
the ``collapse'' of a wave function, at the core of the Copenhagen interpretation of quantum mechanics, is not a
unitary process, for such evolution has no inverse -- one cannot un-collapse a collapsed wave function.
The ``many worlds'' interpretation restores unitarity in a non-falsifiable way.
Perhaps there is a similar point of view lurking behind CTCs.
While it has been shown that when causality is sacrificed in interacting field theories,
then one necessarily loses perturbative unitarity~\cite{Friedman:1990xc,Friedman:1992jc,Boulware:1992pm}
(traceable to the fact that the time-ordering assignment in the Feynman propagator is ambiguous on spacetimes with CTCs).
It has also been proposed that just this sacrifice of unitarity be made in a ``generalized quantum
mechanics''~\cite{Hartle:1993sg}.
And again, in the class of CTCs we consider,
paradox considerations, such as unitarity violation, apply only to the gauge-singlet sector.

Even readers who do not
believe in the possibility of time travel may still find aspects of this article of intellectual interest.
The process of exploring time travel
may provide
a glimpse of the ingredients needed to complete Einstein's gravity.
This completion may require a quantized or emergent theory of gravity, and/or higher dimensions, and/or other.
Furthermore, we will propose specific experimental searches for time travel,
and so stay within the realm of falsifiable physics.


\section{A Class of Metrics Admitting Closed Timelike Curves}
\label{sec:metric}
%
%
The success of ADD model inspires us to think about the possibility of constructing viable CTCs
by the aid of extra dimensions. With the criteria of simplicity in mind, we choose a time-independent
metric and invoke only a single compactified
spatial extra dimension.  We consider the following form for the metric
\bea
\label{metric}
d\tau^2= \eta_{ij}dx^{i}dx^{j}+dt^2+ 2 \,g(u)\,dt\,du -h(u)\,du^2\,,
\eea
where $i,\,j=1,\,2,\,3$, $ \eta_{ij}$ is the spatial part of the Minkowskian metric,
and $u$ is the coordinate of a spatial extra dimension.
For convenience, we set the speed of light $c=1$ on the brane throughout the entire article.
As guided by the wisdom from previous proposals of CTCs, such as G\"{o}del's rotating universe \cite{Godel},
we have adopted a non-zero off-diagonal term $dt\,du$ for a viable CTC. Another simplicity of the above 5D metric is
that its 4D counter-part is completely Minkowskian.
The determinant of the metric is
\be
\label{det}
{\rm Det} \equiv \rm{Det}[g_{\mu\nu}] = g^2+h \,.
\ee
A weak constraint arises from the spacelike nature of the $u$ coordinate,
which requires the signature $\rm{Det} > 0 $ for the whole 5D metric.
In turn this requires that $g^2 +h >0 $ for all values of $u$,
i.e. $h(u) > -g^2 (u)$ at all $u$.
We normalize the determinant by requiring the standard Minkowskian metric on the brane,
i.e., $\rm{Det}(u=0) =g^2(0) +h(0)=+1$.

Since we have never observed any extra dimension experimentally, we
assume that it is compactified and has the topology  $S^1$ of a 1-sphere (a circle).
Due to this periodic boundary condition, the point $u+L$ is identified with $u$, where $L$ is the size of
the extra dimension.
We do not specify the compactification scale $L$
of the extra dimension at this point, as it is irrelevant to our construction of the CTCs.
A phenomenologically interesting number is $L \gtrsim 1/\textrm{TeV}$
since this opens the possibility of new effects at the LHC.
We will adopt this choice in the discussion of possible phenomenology in \S\, \ref{sec:pheno}.

In the coordinates $\{x^\mu, u\}$, our compactified metric with an off-diagonal term $g(u)$
is reminiscent of a cylinder rotating in $u$-space, with axis parallel to the brane.
Again, this geometry is reminiscent of G\"odel's construction
or the van Stockum-Tipler construction.
However, their ``rotating cylinder'' in the usual 4D spacetime is here replaced
with an extra dimension having a compactified $S^1$ topology.
In our case as well as theirs, the metric is stationary but not static,
containing a nonzero off-diagonal term involving both time and space components.

The elements of the metric tensor must reflect the symmetry of the compactified dimension,
i.e., they must be periodic functions of $u$ with period $L$.
This in turn requires that $g(u)$ and $h(u)$ must have period $L$.
Any function with period $L$ can be expressed in terms of a Fourier series with modes
$\sin(\frac{2\pi n\,u}{L})$ and $\cos(\frac{2\pi n\,u}{L})$, $n=0,1,2,\dots$\
Expanded in Fourier modes, the general metric function $g(u)$ is
\be
\label{genmetric}
g(u)=g_0+A -\sum_{n=1}^\infty \left\{
   a_n\,\cos\left(\,\frac{2\pi\,n\,u}{L}\,\right)+b_n\,\sin\left(\,\frac{2\pi\,n\,u}{L}\,\right)
   \right\}\,,
\ee
where $g(0)=g_0$ and $A\equiv \sum_{n=1}^\infty a_n$ are constants.
An analogous expansion can be written down for the metric function $h(u)$,
but in what follows we will not need it.

Below we will demonstrate that the 5D metric we have constructed is sufficient to admit CTCs.
It is worth mentioning that our 5D metric is easily embeddable in further extra dimensions.

\section{Geodesic Equations and their Solutions}
\label{sec:geodesics}

On the brane, the metric in Eq. (\ref{metric}) is completely Minkowskian.
Accordingly, the geodesic equations of motion (eom's) along the brane are simply a vanishing proper acceleration
$\ddot{\vec r}=0$, with dot-derivative denoting differentiation with respect to the proper time $\tau$. Thus,
\beq{dotr}
\dot{\vec r}=\dot{\vec r}_0\,,~~{\rm or}~~~{\vec r}={\vec r}_0\,\tau\,.
\eeq
The geodesic equations for time and for the bulk direction are more interesting.
Since the metric is time-independent (``stationary''), there is a timelike Killing vector with
an associated conserved quantity; the quantity is
\be
\label{timeconst}
\dot{t}+g(u)\,\dot{u}=\gamma_0+g_0\,\dot{u}_0\,,
\ee
where on the right-handed side, we have written the constant in an initial-value form. The initial value of
$\dot{t}$, on the brane, is just the boost factor $\gamma_0$.
From this conserved quantity, we may already deduce that time will run backwards,
equivalently, that $\dot{t}<0$, if $g(u)\,\dot{u}>\gamma_0+\dot{u}_0\,g_0$ is allowed by the remaining geodesic equation.
The remaining geodesic equation involving the bulk coordinate $u$ is
\be
\label{2nd_geodesic}
2\,(g\,\ddot{t} - h\,\ddot{u}) - h'\,\dot{u}^2=0\,,
\ee
where the superscript ``prime" denotes differentiation with respect to $u$.

Taking the dot-derivative of Eq.~(\ref{timeconst}),  we may separately eliminate $\ddot{t}$ and
$\ddot{u}$ from Eq.~(\ref{2nd_geodesic}) to rewrite Eqs.~(\ref{timeconst}) and (\ref{2nd_geodesic})
as
\bea
\label{geodesic_eqns}
\ddot{t}(\tau) &=&  \frac{1}{2} \,\frac{-2g' h+gh'}{g^2+h}\,\dot{u}^2 \,,\\
\ddot{u}(\tau) &=& -\frac{1}{2} \,\frac{2gg'+ h'}{g^2+h} \,\dot{u}^2=-\frac{1}{2} \,\ln' (g^2+h)\,\dot{u}^2 \,.
\eea
The latter geodesic equation is readily solved with the substitution
$\xi\equiv\dot{u}$, which implies that $\ddot{u}=\dot\xi=(d\xi/du)(du/d\tau)=\xi(u)(d\xi/du)$.
Let us choose the initial conditions to be that at $\tau=0$, we have $u=0$.
The solutions for $\dot{u}$ and $u$ are
\be
\label{dotu_soln}
\dot{u}(u)=\frac{\dot{u}_0}{\sqrt{g^2(u)+h(u)}}\,,
\ee
and
\be
\label{u_soln}
\int^{u(\tau)}_0 du\,\sqrt{g^2+h} = \dot{u}_0\,\tau\,,
\ee
the latter being an implicit solution for $u(\tau)$. Having solved explicitly for $\dot{u}(u)$ in Eq. (\ref{dotu_soln}),
we may substitute it into the first of Eq. (\ref{geodesic_eqns}) to gain an equation for $t(u)$.
Alternatively, we may solve the implicit equation in Eq. (\ref{u_soln}) for $u(\tau)$,
and substitute it into Eq. (\ref{timeconst}) to get
\bea
\label{ttau}
t(\tau)= (\gamma_0+g(0)\,\dot{u}_0)\,\tau -\int^{u(\tau)}\, du\;g(u)\,.
\eea

The geodesic equations~ Eqs. \eqref{dotu_soln} and~\eqref{u_soln} depend on ${\rm Det}=g^2+h$
but not on $g$ or $h$ individually.
It therefore proves to be simple and fruitful to fix the determinant to
\be
\label{parameterize}
\Det(u)=g^2(u)+h(u) = 1 \,,\quad\ \forall\ u\,.
\ee
We do so.
With this choice, one readily obtains the eom ${\ddot u}=0$, which implies the solutions
\bea
\label{proper_velocity}
\dot{u}(\tau) &=& \dot{u}_0\,,\quad{\rm and}\\
\label{u}
u(\tau)&=& \dot{u}_0\, \tau\,,~~~~~~ (\,\textrm{mod}~ L\,)\,.
\eea
In analogy to the historical CTCs arising from metrics containing rotation,
we will call the geodesic solutions with positive $\dot{u}_0$ ``co-rotating'',
and solutions with negative $\dot{u}_0$ ``counter-rotating''.
So a co-rotating (counter-rotating) particle begins its trajectory with
positive (negative) $\dot{u}_0$.

We note already at this point the possibility for periodic travel in the $u$-direction with negative time.
From Eqs.~\rf{timeconst} and \rf{proper_velocity}, we have
\beq{timeconst2}
\dot{t}=\gamma_0 -(g(u)-g_0)\,\dot{u}_0\,,
\eeq
and its value averaged over the periodic path of length $L$
\beq{ave_timeconst}
\bar{\dot{t}}=\frac{1}{L}\int_0^L du \;\dot{t} = \gamma_0-(\bar{g}-g_0)\,\dot{u}_0\,,
\eeq
where
\beq{gbar}
\bar{g} = \frac{1}{L}\,\int_0^L\;g(u)\;du = g_0+A \,.
\eeq
is the average value of the metric element along the compact extra dimension.
The latter equality follows immediately from Eq.~\rf{genmetric}.
Thus we have
\beq{signAudot}
\bar{\dot{t}}=\gamma_0-A\,\dot{u}_0\,,
\eeq
which can be negative only if $A$ and $\dot{u}_0$ have the same sign.

Apparently, closing the path in negative time will depend on the relation between the mean value
$\bar{g}$ and the value of the element on the brane $g_0$,
and on the relation between the velocities of the particle along the brane and along the bulk,
characterized by $\gamma_0$ and $\dot{u}_0$.
In the next subsections we examine this possibility in detail.

\subsection{The CTC Possibility}
\label{subsec:CTCpossibility}
By definition, a CTC is a geodesic that returns a particle to the same space coordinates from which it left,
with an arrival time before it left.
The ``closed'' condition of the CTC can be satisfied easily in our metric
due to the $S^1$ topology of the extra dimension.
Namely, if a particle created on the brane propagates into the extra dimension,
it will necessarily come back to $u=0$ due to the periodic boundary condition.\footnote
{
The geodesic equations for travel along the brane are
trivially just constancy of the three-vector part ${\dot{\vec r}}$ of the four-velocity.
Added to the geodesic solution for $u(\tau)$, one gets a constant translation of the
circle $S^1$ along the brane, resulting in a helical motion which periodically intersects the brane
(see \S~\ref{sec:WLsinglets}).
}
The other condition for a CTC, the ``timelike'' condition,
is that the time elapsed during the particle's return path
as measured by an observer sitting at the initial space coordinates is negative.
To ascertain the time of travel, and its sign, we proceed to solve for $t(u)$.
As indicated by Eq. \eqref{ttau}, to do so we need to specify $g(u)$.\footnote
{
Once $g(u)$ is specified,
$h(u)$ is given by $h(u)=1-g^2(u)$, due to the choice made in~ Eq. \eqref{parameterize}.
In particular, the periodicity imposed in $g(u)$ now automatically
ensures that $h(u)$ is periodic, too.
}
Our Fourier expansion of the general compactified metric function (Eq.~\rf{genmetric})
is sufficient for this task.

Our remaining task is to determine $t(u)$ and see if it can be negative.
From Eqs.~\eqref{ttau} and \eqref{u} we have
\beq{intermed}
t(u)=\left( g_0 +\frac{1}{\beta_0}\right)\,u - \int_0^u du\,g(u)\,,
\eeq
where we find it useful to define the symbol
\beq{betazero}
\beta_0=\frac{{\dot u}_0}{\gamma_0}=\left(\frac{du}{dt}\right)_0\,,
\eeq
for the initial velocity of the particle in $u$-direction as would
be measured by a stationary observer on the brane.
For the co-rotating particle, $\beta_0>0$,
while for the counter-rotating particle, $\beta_0<0$.
Performing the integral over Eq.~\rf{genmetric} as prescribed in~ Eq. \rf{intermed}, we arrive at
\beq{tu}
t(u)=\left(\frac{1}{\beta_0}-A\right)\:u
   + \left(\frac{L}{2\pi}\right)\sum_{n=1}^\infty \left(\frac{1}{n}\right)
   \left\{
     a_n\,\sin\left(\frac{2\pi\,n\,u}{L}\right) + b_n\,\left[1-\cos\left(\frac{2\pi\,n\,u}{L}\right)\right]
   \right\}\,.
\eeq
%

\begin{figure*}[ht]
\includegraphics[width=8.5cm]{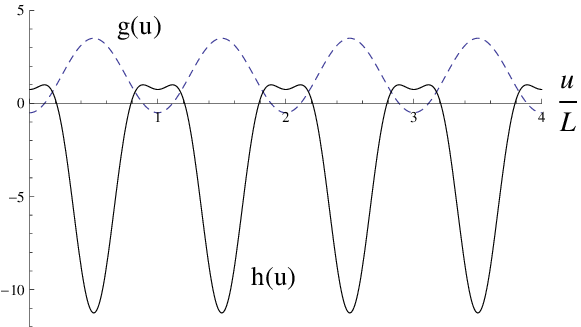}
\caption{
$g(u)$ (dashed) and $h(u)=1-g^2(u)$ (solid) versus $u/L$,
for parameter choices
$g_0=-0.5$, $a_1=A=2$, and $a_{n\neq 1}=b_n=0$
\label{fig:g_and_h}}
\end{figure*}
\begin{figure*}[ht]
\includegraphics[width=8.5cm]{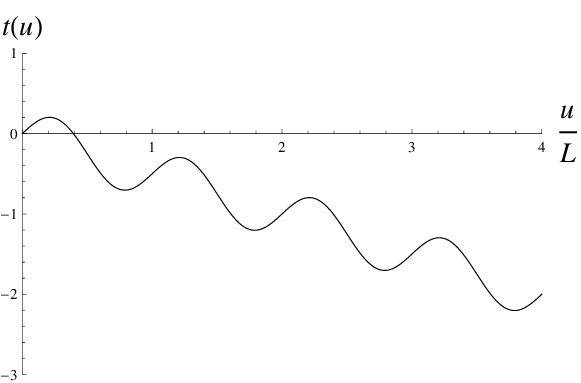}
\caption{
$t(u)$ versus $u/L$, for the same parameter choices as in Fig.~\ref{fig:g_and_h},
and with $\beta_0 \equiv \left( \frac{du}{dt} \right)_0 =2/3$.
\label{fig:t_vs_u}}
\end{figure*}

Due to the $S^1$ topology of the compactified extra dimension, the particle returns to the brane at
$u=\pm NL$,\, for\, integer $N>0$.
The plus (minus) sign holds for a co-rotating (counter-rotating) particle.
Physically, $N$ counts the number of times that the particle has traversed the
compactified extra dimension.
When the particle crosses the brane for the $N^{th}$~time,
the time as measured by a stationary clock on the brane is
\beq{tNL}
t_N\equiv t(u=\pm NL)=\pm \left( \frac{1}{\beta_0}-A\right)\,NL\,.
\eeq
This crossing time depends on the Fourier modes only through $A=\sum_{n=1} a_n$,
and is independent of the $b_n$.
Thus, the potential for a CTC lies in the cosine modes of the metric function $g(u)$,
and not in the sine modes.\footnote
{
This leads to a simple, necessary but not sufficient condition on the metric function for
the existence of a CTC:
for the ${\rm Det}=1$ metric, $g_0$ must differ from $g(\frac{L}{2})$.
}

\subsection{The CTC Realized -- Negative Time Travel}
\label{subsec:CTCrealized}
A viable CTC is realized only if \,$t_N <0$, i.e. $\pm N(\frac{1}{\beta_0}-A)<0$.
For a co-rotating particle ($\beta_0>0$ and positive signature),
a viable CTC requires the conditions\footnote
{
Note that we here assume that $|\beta_0| < 1$ when the particle is created on the brane.
We are allowed to make this assumption because our metric class
will not need superluminal speeds to realize the CTCs.
}
\beq{Acondition}
A > \frac{1}{\beta_0}\, > \,1 \,.
\eeq
On the other hand, for a counter-rotating particle ($\beta_0 <0$ and negative signature),
we require
\beq{Acondition2}
A,\,\beta_0 < 0 \quad{\rm and}\quad  |A|\,>\,\left|\frac{1}{\beta_0}\right|\,>\,1
\eeq
to realize the CTC.
In either case, co-rotating or counter-rotating particles,
the CTC conditions require that sign($A$) be the same as sign($\beta_0$).
Nature chooses the constant $A$ with a definite sign, and so
the CTC conditions for co-rotating and counter-rotating particles
are incompatible.
For definiteness in what follows,
we will assume that it is the co-rotating particles which may traverse the CTC
and not the counter-rotating particles, i.e., that
$A>\frac{1}{\beta_0}\,>\,1$ holds for some $\beta_0$.
The counter-rotating particles of course exist, but they move forward in time.

We note that the negative time of the CTC scales linearly with
the number of times $N$ that the particle traverses the compact $u$-dimension.
The temporal period of this march backwards in time is $\left|\frac{1}{\beta_0}-A\right| L$, with
the natural time-scale being $L/c\sim 10\,(L/{\rm mm})$~picoseconds.

We next give a useful analysis of world-line slopes derived from our metric,
and their connection to time travel.
Such an analysis can offer considerable insight into negative-time physics.

\section{Light-cone/World-line Slope Analysis}
\label{sec:light-cone}
%
A light-cone analysis of the metric, in a fashion similar to the one in~\cite{Tom}, is illuminating.
Here we will make the slight generalization to world-lines of massive particles rather than light-cones of
massless particles.

It is required for the existence of CTCs that the world-line tips so that
evolution in the $u$-direction occurs with backward evolution in time $t$ as measured from the brane.
Let $\tau$ be the proper time of the massive particle in consideration.
When written in terms of the slopes $s\equiv dt/du$ for the world-line in the $\pm u$-directions,
the line element becomes
\beq{slopes1}
\frac{d\tau^2}{du^2}= \frac{1}{{\dot u}^2(u)}=s^2 + 2s\,g(u) -h(u)\,,
\eeq
and we have neglected possible nonzero \,$d{\vec r}/dt$ since it does not affect this discussion.
The solutions of this quadratic equation are the two slopes for the co-rotating and counter-rotating
world-lines:\footnote
{One obtains the (massless particle) light-cone results by setting $\frac{1}{{\dot u}^2}$
(proportional to $d\tau^2$) to zero.
}
\be
\label{slopes2}
s_\pm (u) = -g(u)\pm\sqrt{g^2(u)+h(u)+\frac{1}{\dot{u}^2(u)}}\,.
\ee
To ascertain the assignment of the two world-lines to the co-rotating and counter-rotating
particles, we note that by definition, $s=\frac{dt}{du}$, so $s(u=0)=\frac{\gamma_0}{{\dot u}_0}$.
Thus, sign($s(0)$)= sign($\dot{u}_0$), i.e., $s_+$ is the world-line for the ${\dot u}>0$
co-rotating particle, and $s_- $ is the world-line for the ${\dot u}<0$ counter-rotating particle.

Equivalent to Eq.~\rf{slopes2} are
\be
\label{sumNproduct}
s_-(u) \,+ \,s_+(u) = -2\,g(u)
     \quad {\rm and\ }
s_-(u)\;s_+ (u) = -\left(\,h(u)+\frac{1}{\dot{u}^2(u)}\,\right)\,.
\ee
If the $u$-direction were not warped, we would expect the Minkowskian result
$s_- \,+\, s_+ = 0$ and $s_-\;s_+ = -(1+{\dot u}^{-2})$.
Instead, here we have
$s_{-}(0)\, +\, s_{+}(0)= -2\,g_0 $ and $s_-(0)\;s_+ (0) = -h_0 +\dot{u}_0^{-2}$, where $h_0\equiv h(0)$.
So for $g_0\ne 0$ \,and/or \,$h_0 \neq 1$,
we have a warped dimension beginning already at the brane slice $u=0$.
We maintain Minkowskian-like behavior at $u=0$ by choosing $h_0$ to be non-negative.

For the world-line to tip into the negative $t$ region, its slope must pass through zero.
This requires the product $s_- (u)\; s_+ (u)$ and hence $h(u)+\dot{u}^{-2}$ to pass through zero.
We label the value of $u$ where this happens as $\ustar$.
Thus, $h(\ustar)=-{\dot\ustar}^{-2}$.
Positivity of the metric determinant $g^2+h$ at all $u$ then demands
at $\ustar$ where $h(\ustar)+\dot{\ustar}^{-2}$ vanishes,
that $g^2(\ustar)> {\dot\ustar}^{-2} \neq 0$.

Next we implement our simplifying assumption that $g^2+h=1$
and its concomitant result ${\dot u}={\dot u}_0,\ \forall\,u$~
(Eq.~\eqref{proper_velocity}).
As a result,
(i) the condition $h_0\ge 0$ in turn implies that $|g_0|\le 1$;
(ii) we have $g^2(\ustar)=1-h(\ustar)=1+\dot{u}_0^{-2}$,
so the condition $g^2(\ustar)\neq 0$ is automatically satisfied.

Importantly, time will turn negative if $g(u)$ rises from its value $|g_0|\le 1$ on the brane
to above $\sqrt{1+\dot{u}_0^{-2}}$.
Such behavior of $g(u)$ is easy to accommodate with a metric function
as general as Eq.~(\ref{genmetric}).
In Fig.~\rf{fig:g_and_h} we show
sample curves for $g(u)$ and $h(u)=1-g^2(u)$.
A quick inspection of the $g(u)$ shown in this figure
convinces one that even a simple metric function
can accommodate time travel. In Fig.~\rf{fig:t_vs_u}, we show explicitly the accumulation of negative
time as the particle travels around and around the extra dimension.

One lesson learned from this slope analysis is that only co-rotating or only counter-rotating
particles, but not both, may experience CTCs.
This is because only {\sl one} edge of the lightcone tips below the horizontal axis
into the negative-time half-plane.
The development of co-rotating and counter-rotating geodesics in the previous section
is consistent with this lesson. Another lesson learned is that CTCs may exist for large
${\dot u}_0$, but not for small ${\dot u}_0$;
i.e., there may exist a critical $({\dot u}_0)_{\rm min}$ such that CTCs
exist for ${\dot u}_0> ({\dot u}_0)_{\rm min}$,
but not for ${\dot u}_0< ({\dot u}_0)_{\rm min}$.
Finally, we remark that the slope analysis presented here
may be derived from a more general covariant analysis.
The connection is shown in Appendix~\rf{app:GuthTime}.

\section{Conditions on Metric Parameters that Allow CTCs}
\label{sec:conditions}
So far, the conditions on the parameters of the metric that must be obeyed if CTCs are admissable,
are two in number: From Eqns.~\rf{Acondition} and \eqref{Acondition2} that $|A| = | \bar{g}-g_0 | > 1$,
and from the previous section that $|g_0| \leq 1$.
Together, these two conditions imply that the sign of $\bar{g}=g_0+A$, i.e. $sign(\bar{g})$, is the same as $sign(A)$.
In \S\rf{subsec:CTCrealized} it was inferred that CTCs require that $sign(A)=sign(\beta_0)$, which in turn has the sign of
$\dot{u}_0=\beta_0\gamma_0$.  Thus we have the inference from the two stated conditions that $sign(\bar{g})=sign(\dot{u}_0)$.
We will make use of this inference below.

There is a stronger condition to be imposed.
It was shown in a simple way that $|\bar{g}| > \bar{D}$ must hold in our metric in order that massless particles
travel at the speed of light when  global ``diagonalized'' coordinates $\tilde{t}$ and $\tilde{u}$ are employed~\cite{Gielen}.
$D$ is defined as the square root of the metric's determinant ${\rm Det}=g^2+h$,
and $\bar{D}$ is its value averaged over the compact extra dimension.
In our work, we set Det everywhere equal to unity, its Minkowski value on the brane.
Thus, the condition becomes $| \bar{g} | \ge 1$ for us.
The coordinates $\tilde{t}$ and $\tilde{u}$ which diagonalize the metric into Minkowski form
are defined in differential form in Eq.~\rf{diffs}, and in integrated form in Eqns.~\rf{ubar} and~\rf{tbar}.
These coordinates are pathological in a sense to be discussed in \S~\rf{sec:compared2spinning}),
but useful for theoretical proofs.
However, this time is not a variable that would register on a clock of an (LHC) experimenter.

Here we show how this constraint equation may be derived using the standard $t$ and $u$ coordinates.
Three constants of geodesic motion have been identified in Eqs.~\rf{dotr}, \rf{timeconst}, and~\rf{proper_velocity}
as $\dot{\vec{r}}$,\, $\dot{t}+g(u)\dot{u}$, and $\dot{u}$, respectively.
The first two constants are inevitable results of the metric depending only on the coordinate $u$,
while the latter constant results when our ansatz ${\rm Det}=1$ is implemented.
Equivalent to any one of the geodesic constants of the motion is the ``first integral'' constructed by
dividing the line element in Eq.~\rf{metric} by $d\tau^2$:
\beq{first1}
\zeta=\dot{t}^2+2\,g(u)\,\dot{t}\,\dot{u}-h(u)\,{\dot{u}}^2-{\dot{\vec{r}}}\;^2\,,
\eeq
where $\zeta=1$ for matter, and zero for photons.
Substituting in the first two constants of motion, and rearranging the right-hand terms a bit,
one gets
\beq{first2}
\zeta=\left (\,\dot{t}+g(u)\,\dot{u}\,\right)^2 - \left(\,{\rm Det}(u)\,{\dot{u}}^2 +{ \dot{\vec{r}}}\;^2\,\right) \,.
\eeq
Then, making use of Eq.~(3.6),
we arrive at
\beq{first3}
\zeta=\left(\, \dot{t}+\left(\,\frac{g(u)}{\sqrt{\rm Det}}\,\dot{u}_0\,\right)\,\right)^2 - \left(\,{\dot{u}_0}^2 +{ \dot{\vec{r}}}\;^2\,\right) \;.
\eeq
Rearranging terms again and then taking the square root gives
\beq{first4}
\left| \, \dot{t}+\left(\,\frac{g(u)}{\sqrt{\rm Det}}\,\right)\,\dot{u}_0 \, \right| =\sqrt{ \zeta+ {\dot{u}_0}^2+{ \dot{\vec{r}}}\;^2 }\,.
\eeq
Next, we take the average over the extra-dimensional transit to get
\beq{first5}
\left| \, \langle\dot{t}\rangle+\left\langle\frac{g}{\sqrt{\rm Det}}\,\right\rangle\,\dot{u}_0 \,\right| =\sqrt{ \zeta+ {\dot{u}_0}^2+{ \dot{\vec{r}}}\;^2 }\,.
\eeq

Since $sign(\bar{g})=sign(\dot{u}_0)$, and $\bar{D}>0$ everywhere on the geodesic,
we make the conservative assumption that
$\left\langle\frac{g}{\sqrt{\rm Det}}\,\right\rangle\,\dot{u}_0 >0$.
Then, noting that $\langle\dot{t}\rangle=0$ for a closed null curve and $\langle\dot{t}\rangle < 0$ for a pre-arriving particle,
and rewriting  $\dot{u}_0$ in its equivalent form $\beta_0 \gamma_0$
and $\dot{\vec{r}}_0$ in its equivalent form $\vec{v_0} \gamma_0$,
we arrive at the final expression for the necessary and sufficient condition for a CTC or a pre-arrival to occur:
\beq{first6}
\left| \left\langle\frac{g}{\sqrt{\rm Det}}\right\rangle\right| \, \ge
\sqrt{ 1+\frac{
(\zeta/\gamma_0^2)+\vec{v}_0\,^2}{\beta_0^2} } \;.
\eeq
(We do not consider here the alternate mathematical  solution with very large, negative $\langle\dot{t}\rangle$;
this solution does not connect continuously to the $\langle\dot{t}\rangle=0$ CTC condition.)

The particle's boost factor $\gamma_0$ may greatly exceed unity,
and the particle's velocity $\vec{v}_0$ along the brane may be zero.
Thus, we may write the necessary but not sufficient condition for a CTC or pre-arrival as simply
\beq{first7}
\left| \left\langle\frac{g}{\sqrt{\rm Det}}\right\rangle \right| > 1 \,.
\eeq
For a photon (with $\zeta=0$) but not for massive particles such as Higgs singlet KK modes (with $\zeta=1$) of interest to us,
the inequality becomes ``$\ge$''.
Thus, with Det taken to be unity, we arrive at the constraint $|\bar{g}| \ge 1$  for photons
(in agreement with result obtained in diagonalized coordinates~\cite{Gielen}.)

The ultimate conditions on the metric which guarantee CTC solutions are now three in number, and simple.
They are
(i) $|\bar{g}| > 1$~for massive particles, with $sign(\dot{u}_0)=sign(\bar{g})$ (taken to be positive for CTCs, by convention),
as just derived;
(ii) $|g_0| \leq 1$, as derived in \S \rf{sec:light-cone};
and (iii) $|A| = | \bar{g}-g_0 | > 1$ from Eqs.~\rf{Acondition} and \eqref{Acondition2}.
In Fig.~\rf{fig:g0_vs_A}
we display the shaded region in the $g_0$-$\bar{g}$~plane that allows CTCs.
The allowed positive values of $\bar{g}$ extend to $+\infty$.
We note that the parameters used in Figs.~\rf{fig:g_and_h} and~\rf{fig:t_vs_u}
were chosen to respect these three constraints.

\begin{figure*}[t!]
\includegraphics[height=5cm,width=10cm]{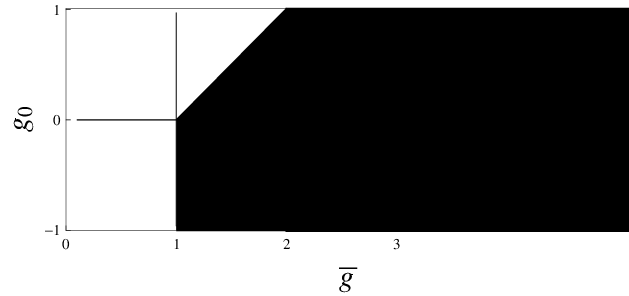}
\caption{The region (shaded) in the $g_0$-$\bar{g}$~plane for which CTCs are possible.
\label{fig:g0_vs_A}}
\end{figure*}

\section{Compactified 5D CTCs Compared/Contrasted with 4D Spinning String}
\label{sec:compared2spinning}
Before turning to phenomenological considerations of particles traversing CTCs,
we wish to show that the compactified 5D metric admitting CTC geodesics
is devoid of pathologies that plague similar 4D metrics.
This class of 5D metrics resembles in some ways the well-studied metric
which describes a spinning cosmic string.

\subsection{4D Spinning String(s)}
\label{4Dstring}
The metric for the 4D spinning string is~\cite{DJtH-Annals,DJ-Feinberg}:
\beq{spinstr}
d\tau^2_{\rm\stackrel{spinning}{string}}= (dt+ 4\,G\,J \,d\theta)^2-dr^2-(1-4\,G\,m)^2\,r^2\,d\theta^2-dz^2\,,
\eeq
where $G$ is Newton's constant, and $J$ and $m$ are the angular momentum and mass per unit length
of the cosmic string, respectively.
In three spacetime dimensions, the Weyl tensor vanishes and any source-free region is flat.
This means that in the region outside of the string,
local Minkowski coordinates may be extended to cover the whole region.
Specifically, by  changing the coordinates in Eq.~\rf{spinstr}
to $\tilde{t} = t+ 4\,G\,J \,\theta $ and $\varphi =(1-4\,G\,m)\,\theta $,
the metric appears Minkowskian, with the conformal factor being unity.
As with $\theta$, the new angular coordinate $\varphi$ is periodic, subject to the identification
$\varphi \sim \varphi + 2\,\pi- 8\,\pi \,G\,m$.
There is a well-known wedge $\Delta\varphi=8\,\pi \,G\,m$
removed from the plane, to form a cone.
However, although this transformation appears to be elegant, in fact
the new time $\tilde{t}$ is a pathological linear combination of a compact variable $\theta$
and a non-compact variable $t$:
for fixed $\theta$ (or $\varphi$) one expects $\tilde{t}$ to be a smooth and continuous variable,
while for fixed $t$, one expects the identification $\tilde{t} \sim \tilde{t}+ 8\,\pi\,G\,J$
in order to avoid a ``jump'' in the new variable.
In effect, the singularity at $g_{\theta\theta}=0$, i.e., at $r=4GJ/(1-4Gm)$,
is encoded in the new but pathological coordinate $\tilde{t}$~\cite{DJtH-Annals}.

The form of our metric in the $(t,\ u)$-plane,
viz.\
\beq{ut_form}
d\tau^2=dt^2+2g(u)\,dt\,du - h(u)\,du^2 = (dt+g(u)\,du)^2 - (g^2(u)+h(u))\,du^2\,,
\eeq
has similarities with the spinning string metric.
Analogously, we may define new exact differentials
\beq{diffs}
d\bar{t}\equiv dt+g(u)\,du\,,\quad{\rm and}\quad d\bar{u}\equiv \sqrt{g^2(u)+h(u)}\,du\,,
\eeq
which puts our metric into a diagonal ``Minkowski'' form:
\bea
\label{metric2}
d\tau^2 &=& \eta_{ij}\,dx^{i}dx^{j}+ d\bar{t}^2 - d\bar{u}^2\,.
\eea
Being locally Minkowskian everywhere, the entire 5D spacetime is therefore flat.
This accords with the theorem which states that
any two-dimensional (pseudo) Riemannian metric
(whether or not in a source-free region) --- here, the $(u,t)$ submanifold ---
is conformal to a Minkowskian metric.
In our case, the geometry is $M^4 \times S^1$, which is not only conformally flat,
but flat period (the conformal factor $\Omega(u,t)$ is unity).
However, the topology of our 5D space, like that of the spinning string, is non-trivial.
The new variable $\bar{t}$, defined by $d\bar{t}=dt+g(u)\,du$, is ill-defined globally,
being a pathological mixture of a compact ($u$) and a non-compact ($t$) coordinate.\footnote
{
This nontrivial transformation effectively defines a new time $\bar{t}$ measured in the frame that
``co-rotates'' with the circle $S^1$.
The integrated, global version of these new coordinates is
\beq{ubar}
{\bar u}=\int_0^{\bar u} d\bar{u}=\int_0^u du\,\sqrt{\Det(u)} =u\,,\quad{\rm since\ }\Det=1\,,
\eeq
\bea
\label{tbar}
\bar{t} &=& \int_0^{\bar{t}} d\bar t = \int_0^t dt +\int_0^{u(t)} g(u)\;du = t+ \int_0^{u(t)}\,du\;g(u)\,, \nonumber \\
&=& t+(g_0+A)\:u-\left(\frac{L}{2\pi}\right)\sum_{n=1}^\infty\left(\frac{1}{n}\right)
        \left\{a_n\sin \left(\frac{2\pi\,n\,u}{L}\right)+b_n
	   \left[ 1-\cos\left(\frac{2\pi\,n\,u}{L}\right) \right]
   \right\}
\,.
\eea
}
Thus, the parallel between the metric for a spinning cosmic string
and our metric is clear.

We chose to not exploit the ``Minkowski'' coordinates because the new time $\tilde{t}$ is  necessarily a mixture of the continuous variable $t$ and the compact variable $u$. As the title of our paper suggests, our focus is whether causality-violation may be observable at experimental facilities such as the LHC. Our answer is affirmative, as it was in the original version of our manuscript.

We remark that time as measured by an observer (or experiment) on our brane is just given by
the coordinate variable $t$.
This is seen by constructing the induced 4D metric.
The constraint equation reducing the 5D metric to the induced 4D metric is simply
$u(x^\mu)=0$.  Taking the differential yields $du=0$.
Inputting the latter result into the 5D metric of Eq.~\rf{metric}
induces the standard 4D Minkowski metric.

Next we investigate whether or not our metric suffers
from fundamental problems commonly found in proposed 4D~metrics with CTCs.

\subsection{4D Spinning String Pathologies}
\label{subsec:4Dpathology}
Deser, Jackiw, and 't~Hooft~\cite{DJtH-Annals} showed that the metric for the spinning string admits CTCs.
This metric has been criticized, by themselves and others, for the singular definition of spin
that occurs as one approaches the string's center at $r=0$.
With our metric, there is no ``$r=0$'' in $u$-space --
the ``center'' of periodic $u$-space is simply not part of spacetime.

An improved CTC was proposed by Gott, making use of a pair of cosmic strings with a relative velocity --
spin angular-momentum of a single string is replaced with an orbital angular-momentum of the two-string system.
Each of the cosmic strings
is assumed to be infinitely long and hence translationally invariant along the $z$
direction.  This invariance allows one to freeze the $z$ coordinate,
thereby reducing the problem to an effective (2+1)~dimensional spacetime with two particles at the sites
of the two string piercings.
A CTC is found for a geodesic encircling the piercings and crossing between them.

The non-trivial topology associated with Gott's spacetime
leads to non-linear energy-momentum addition rules.
What has been found is that while each of the spinning cosmic strings carries an acceptable
timelike energy-momentum vector, the two-string center-of-mass energy-momentum vector
is spacelike or tachyonic~\cite{Hooft,Shore}. Furthermore, it has been shown that in
an open universe, it would take an infinite amount of energy to form Gott's CTC~\cite{Carroll}.

Blue-shifting of the string energy is another argument against the stability of Gott's CTC~\cite{Tye}.
Through each CTC cycle, a particle gets blue-shifted~\cite{Hawking,Carroll}.
Since the particle can traverse the CTC an infinite number of times, it can be infinitely blue-shifted,
while the time elapsed is negative.
Total energy is conserved, and so the energy of the pair of cosmic strings
is infinitely dissipated even before the particle enters the CTC for the first time.
Hence, no CTC can be formed in the first place.

\subsection{Pathology-Free Compactified 5D CTCs}
\label{subsec:nopathology}
Our class of 5D metrics seems to be unburdened by the pathologies~\cite{Hooft,Shore} described immediately above.
One readily finds that all the components of the 5D~Einstein tensor as determined by the metric in Eq. (\ref{metric})
are identically zero.
Therefore, by the Einstein equation, the energy-momentum tensor $T_{AB}$ also vanishes.
This implies that our class of metrics with CTCs automatically satisfies all of the
standard energy conditions.\footnote
{The standard energy conditions are,
for any null vectors $l^{A}$ and timelike vectors ${t^{A}}$,
\bea
\textrm{Null Energy Condition (NEC):}~~~ && T_{AB}\, l^{A}\,l^{B} \geq 0 \,, \\
\textrm{Weak Energy Condition (WEC):} ~~~ && T_{AB}\, t^{A}\,t^{B} \geq 0 \,, \\
\textrm{Strong Energy Condition (SEC):} ~~~ && T_{AB}\, t^{A}\,t^{B} \geq \frac{1}{2}\,
     T^{A}_{A}\, t^{B}\,t_{B} \,,   \\
\textrm{Dominant Energy Condition (DEC):} ~~~ && T_{AB}\, t^{A}\,t^{B} \geq 0~~
     \textrm{and} ~~T_{AB}\, T^{B}_{C} \,t^{A}\,t^{C} \leq 0\,.
\eea}
Hawking has conjectured~\cite{Hawking} that Nature universally protects chronology (causality)
with applications of physical laws.  The protection hides in the details.
Hawking has proved his conjecture for the case where the proposed CTC violates the weak energy condition (WEC).
Such WEC violations typically involve exotic materials having negative energy--density.
Our CTCs do not violate the WEC, and in fact do not require matter at all.
With our metric, it is the
compactified extra dimension with the $S^1$ topology
rather than an exotic matter/energy distribution that enables CTCs.
In the realm of energy conditions, our metric contrasts again with Gott's case of two moving strings.
Gott's situation does not violate the WEC,
but it can be shown that the tachyonic total energy-momentum vector
leads to violation of all the other energy conditions.

Furthermore, particles traversing our CTCs are \emph{not} blue-shifted,
unlike the particles traversing Gott's CTC.
This can be seen as follows.
One defines the contravariant momentum in the usual way, as
\beq{PupA}
p^A \equiv m\,(\dot{t},\,\dot{\vec{r}},\,\dot{u})\,,
\eeq
where $m$ is the mass of the particle.
Then the covariant five-momentum is
\beq{PdnA}
p_A = G_{AB}\,p^B = m\,(\dot{t}+ g \,\dot{u},\;-\dot{\vec{r}},\;g\,\dot{t}-h\,\dot{u})\,.
\eeq
According to Eq.~\rf{timeconst}, the quantity $p_0= m \,(\dot{t}+ g \,\dot{u})$ is covariantly conserved
along the geodesic on and off the brane.
The conservation is a result of the time-independence of the metric $G_{AB}$.
Consequently, we identify the conserved quantity as the particle energy $E$
and conclude that particles traversing the compactified 5D CTCs are not blue-shifted.
In a more heuristic fashion, one may say that energy conservation on the brane follows from
the absence of an energy source;\, $T_{AB}$ vanishes for our choice of metric class.

In addition to conservation of $p_0$, conservation of
the particle's three-momentum along the brane follows immediately from the eom and
solution Eq.~\rf{dotr}.
The only component of the particle's covariant five-momentum that is not conserved along the geodesic
is \,$p_5=m\,(\, g\,\dot{t}-h\,\dot{u}\,)$.  This quantity may be written as\, $(\,g\,E-m\dot{u}\,)$,
where use has been made of the relation\, $g^2+h =1$.
But even here, the factors $E$ and (from Eq.~\rf{proper_velocity}) $\dot{u}$ are conserved quantities,
and so it is just the factor $g(u)$ that varies along the geodesic.
However, the metric elements including $g(u)$ are periodic in $u=NL$, and return to their brane values
at each brane piercing.
Thus, the particle's entire covariant five-momentum is conserved from the viewpoint of the brane.
We conclude that the possible instability manifested by a particle's blue-shift~\cite{Hawking,Carroll,Tye}
does not occur in our class of 5D~metric, nor do any other kinematic pathologies.
(Conservation and non-conservation of the components of the particle's five-momentum are discussed from
another point of view in Appendix~\rf{app:p5}.)


In summary, we have just shown that while our class of 5D metrics bear some resemblance to
the metric of the 4D spinning string, the class is free from the $r\rarr 0$ pathology of the spinning string,
does not violate the standard energy conditions as does Gott's moving string-pair, and does not present
particle blue-shifts (energy gains) as does Gott's metric.

\section{Stroboscopic World-lines for Higgs Singlets}
\label{sec:WLsinglets}
%
The braneworld model which we have adopted has SM gauge particles trapped on our 3+1
dimensional brane, but gauge-singlet particles are free to roam the bulk as well as the brane.
We are interested in possible discovery of negative time travel at the LHC, sometimes
advertised as a ``Higgs factory''. The time-traveling Higgs singlets can be produced either from the decay of SM Higgs
or through mixing with the SM Higgs.
We discuss Higgs singlet production in~\S~\rf{sec:pheno}.

\subsection{Higgs Singlet Pre/Re--Appearances on the Brane}
\label{subsec:singlet-appears}
The physical paths of the Higgs singlets are the geodesics which we calculated in previous sections.
The geodesic eom's for the four spatial components of the five-velocity are
trivially $\ddot{\vec r}$ and $\ddot u =0$.
Thus, the projection of the particles position onto the brane coordinates is
\beq{branecoord1}
\vec{r}\:(\tau)= \dot{\vec{r}}_{\ 0}\,\tau +\vec{r}_0 = \dot{\vec{r}}_{\ 0}\,\frac{u}{{\dot u}_0} +\vec{r}_0
    = \frac{v_0}{\beta_0}\,u\ {\hat p}_0+\vec{r}_0\,.
\eeq
Here, $\hat{p}_0=\widehat{(\frac{dr}{d\tau})}_0= \widehat{(\frac{dr}{dt})}_0$
is the unit direction vector of the particle's three-momentum as seen by a brane observer,
$v_0$ is the initial speed of the particle along the brane direction\footnote
{
Recall that $\dot\rvec=\frac{d\rvec}{d\tau}$ is a constant of the motion,
but $\vec{v}=\frac{d\rvec}{dt}$ is not, due to the non-trivial relationship between
$\tau$ and $t$.  This non-trivial relationship between proper and coordinate times
is what enables the CTC.
}
\beq{initial-speed}
v_0\equiv\left|\left(\frac{d\rvec}{dt}\right)_0\right|\,,
\eeq
and $\rvec_0$ is the point of origin for the Higgs singlet particle, i.e.,
the primary vertex of the LHC collision.

Of experimental interest is the reappearance of the particle on the brane.
Inserting $u=\pm NL$ into~ Eq. \rf{branecoord1}, one finds that
the particle crosses the brane stroboscopically;
the trajectory lies along a straight line on the brane, but piercing the brane at
regular spatial intervals given by
\beq{branecoord2}
\vec{r}_N = \frac{v_0}{|\beta_0|}\,NL\ \hat{p}_0+\rvec_0\,.
\eeq
We note the geometric relation
$\frac{v_0}{\beta_0}= \frac{|\dot{\rvec}_0|}{\dot{u}_0}= \cot\theta_0$,
where $\theta_0$ is the exit angle of the particle trajectory
relative to the brane direction.

The result in Eq. \rf{branecoord2} for the spatial intervals on the brane
holds for both co-rotating particles and counter-rotating particles.
The distance between successive brane crossings, $L\,v_0/\beta_0$,
is governed by the size $L$ of the compactified dimension.
These discrete spatial intervals are likely too small to be discerned.
However, the Higgs singlet is only observed when it scatters or decays
to produce a final state of high-momenta SM particles.
We expect the decay or scattering rate to be small, so that many bulk orbits are traversed
before the Higgs singlet reveals itself.
We discuss many-orbit trajectories next.

\subsection{Many-Orbit Trajectories and Causality Violations at the LHC}
\label{subsec:LHC}

The coordinate times of the reappearances of the particle on the brane are given
by Eq.~\rf{tNL} as
\beq{tNagain}
t_N =\pm \left( \frac{1}{\beta_0}-A\right)\,NL\,.
\eeq
We have shown earlier that the assumption $A>0$, without loss of generality,
forces co-rotating particles having $A>\frac{1}{\beta_0}>1$ to travel in negative time.
Thus, for the particles traveling on co-rotating geodesics (with $\beta_0 >0$ and positive signature),
their reappearances are in fact pre-appearances!
The time intervals for pre-appearances are
\beq{co-times}
t_N({\rm co{\rm -}rotating})= -\left( A-\frac{1}{\beta_0}\right)\,NL<0\,.
\eeq
For counter-rotating particles (with $\beta_0 <0$ and negative signature),
the time intervals are
\beq{counter-times}
t_N({\rm counter{\rm -}rotating})= \left(A+\left|\frac{1}{\beta_0}\right|\right)\,NL>0\,.
\eeq
The counter-rotating particles reappear on the brane at regular time intervals,
but do not pre-appear.
We note that even for co-rotating and counter-rotating particles
with the same $|\beta_0|$, the magnitudes of their respective time intervals are different;
the co-rotating interval is necessarily shorter.
The mean travel times for co-rotating and counter-rotating particles, respectively, are
\beq{co-meantimes}
\langle t\rangle ({\rm co{\rm -}rotating})=
   -\left( A\langle N\rangle-\left\langle\frac{N}{\beta_0}\right\rangle\right)\,L<0
\eeq
and
\beq{counter-meantimes}
\langle t\rangle ({\rm counter{\rm -}rotating})=
   \left(A\langle N\rangle+\left\langle\frac{N}{|\beta_0|}\right\rangle\right)\,L>0\,.
\eeq
%
In \S\rf{sec:pheno} we will show that these means are very large numbers,
inversely related to the decay/interaction probability of the Higgs singlet.

\subsection{Higgs Singlet Apparent Velocities Along the Brane}
\label{subsec:singlet-velocity}
%
%
%

\begin{figure*}[ht]
\includegraphics[width=9.5cm]{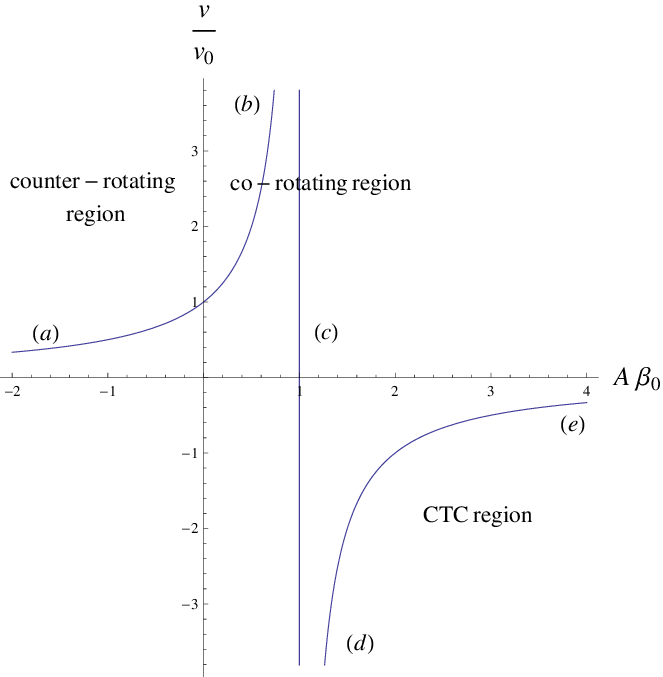}
\caption{
Apparent brane velocity $v$ as fraction of initial brane velocity $v_0$
versus $\beta_0\:A$.  The counter-rotating particle always moves subluminally forward in time,
but the co-rotating particle may move superluminally in either time direction.
Brane velocities are divergent at $\beta_0\:A = 1$, which occurs as the lightcone
crosses the horizontal axis of the spacetime diagram.
For $\beta_0\:A > 1$, the co-rotating geodesic is a CTC.
The regions delineated by (a), (b), (c), (d), (e) map into
the world lines of Fig.~\ref{fig:worldlines} with the same labels.
\label{fig:regions}}
\end{figure*}
\begin{figure*}[ht]
\includegraphics[width=8cm]{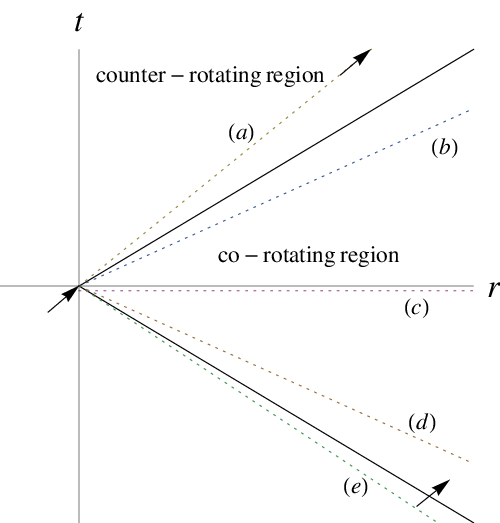}
\caption{
Shown are stroboscopic piercings (dots) of our brane by a returning Higgs singlet.
World lines delineated by (a), (b), (c), (d), (e) correspond to
regions of Fig.~\ref{fig:regions} with the same labels.
In (a), the counter-rotating particle travels forward in brane/coordinate time,
within the forward light-cone.
The co-rotating particle travels outside the brane's forward light-cone.
In (b), the world line is superluminal but moving forward in brane time $t$.
In (c), the world line is horizontal; the particle ``moves'' instantaneously in brane time.
In (d) and (e), the particle travels
 superluminally and subluminally, respectively,
backwards in brane time (signifying a CTC).
\label{fig:worldlines}}
\end{figure*}

We may compute the observable velocity connecting the production site for the Higgs singlet to
the stroboscopic pre-and re-appearances by dividing the particle's apparent travel distance
along the brane, ${\vec r}_N(t)-{\vec r}_0$ given in Eq.~\rf{branecoord2},
by the apparent travel time $t_N$ given in Eq.~\rf{tNagain}.
These are the velocities which an observer on the brane, e.g. an LHC experimenter, would infer from measurement.
For the co-rotating particles with $\beta_0 >0$, we get
\beq{v_coro}
\vec{v}\,({\rm co{\rm -}rotating})=-\,\frac{v_0}{\beta_0\,A-1}\ {\hat p}_0\,.
\eeq
The velocities of the co-rotating particles are negative for $\beta_0 A >1$
because the particles are traveling backwards in time.
For the counter-rotating particles with $\beta_0 <0$, we get
\beq{v_counter}
\vec{v}\,({\rm counter{\rm -}rotating})=\frac{v_0}{|\beta_0|\,A+1}\ {\hat p}_0\,.
\eeq
The velocities of the counter-rotating particles are positive,
as the particles travel forward in time, and subluminal.
For a given exit angle $\theta_0=\cos^{-1}(v_0/\beta_0)$,
the speeds of the co-rotating particles, traveling backwards in time,
generally exceed the speeds of the counter-rotating particles.

We note that the apparent speeds of co-rotating particles can be superluminal in either
forward time ($\beta_0\:A<1$) or backwards time ($\beta_0\:A >1$).
We display the velocities in Fig.~\rf{fig:regions} with a plot of
$v/v_0$ versus the parameter combination $\beta_0 A$.
The particle speed diverges at $\beta_0 A = 1$; the value $\beta_0\:A=1$ corresponds to the slope of the light-cone passing through zero,
an inevitability discussed in \S~\rf{sec:light-cone}.
The region $\beta_0 A > 1$ is the CTC region,
of interest for this article.

In Fig.~\rf{fig:worldlines} we show schematically the world lines on our brane for co-rotating particles
with negative transit times, and for counter-rotating particles with positive~($\beta_0\:A<1$)
and negative~($\beta_0\:A>1$) transit times.

At some point along the world line, during one of the brane piercings,
the Higgs singlet decays or interacts to produce a secondary vertex.
As we show in the next paragraph,
during each brane piercing, the particle's three-momentum is just that missing from the
primary vertex, i.e., three-momentum on the brane is conserved.
The arrows in Fig.~\rf{fig:worldlines} are meant to denote the three-momentum
missing from the primary vertex at the origin, and re-appearing or pre-appearing in
a displaced secondary vertex.
The piercings (dots) of the brane have a $\Delta r\sim$
the brane width~$w$, and a $\Delta t\sim w/v_0$.
We have seen in \S~\rf{subsec:nopathology} that three-momentum (in fact, five-momentum) on the brane is conserved
for time-traveling particles.
Therefore, the particle slopes $\Delta t/\Delta r \sim 1/v_0$ give no indication whether the particles are
negative time-travelers.
Only the pre-appearance of the secondary vertex with respect to the
primary vertex reveals their acausal nature.
Importantly, the secondary and primary vertices are correlated by the conservation of particle three-momentum:
exactly the momentum missing from the primary vertex pre-appears in the secondary vertex.

\section{5D and Effective 4D Field Theory for Time-Traveling Higgs Singlets}
\label{sec:pheno}

In previous sections, we have constructed a class of 5D metrics which admits stable CTC solutions of the classical Einstein equations,
and we have presented the solutions.
Similar to the ADD scenario, we will assume that all the
SM particles are confined to the brane while gauge singlets, such as Higgs singlets, gravitons and sterile neutrinos,
can propagate into the bulk.
In this section, we first construct the 5D Lagrangian for the coupled Higgs singlet--doublet system.
Then we derive the 5D equation of motion (5D Klein-Gordon equation) in our spacetime,
and solve it subject to the compactified $5^{th}$~dimension boundary condition.
From this exercise, there results the interesting energy-momentum dispersion relation.
Next we integrate out the $5^{th}$~dimension to obtain the effective 4D theory.
Finally, we incorporate electroweak (EW) symmetry-breaking to obtain the effective theory
relevant for EW-scale physics.

\subsection{5D Lagrangian for the Coupled Higgs Singlet--Doublet System}
\label{subsec:5D4Dlagrangians}
A simple and economical model involves the Higgs singlet $\phi$ coupling/mixing only with the SM Higgs doublet $H$.
We add the following Lagrangian density $\mathcal{L}^{(5D)}=\mathcal{L}_0 + \mathcal{L}_I$ to the SM:
\bea
\label{L5D_0}
\mathcal{L}_0 &=& \frac{G^{AB}}{2}\,\partial_A \,\phi \,\partial_B \,\phi
- \frac{m^2}{2} \, \phi^2 \,, \\
\label{L5D_I}
\mathcal{L}_I &=& -\frac{\lambda_1}{\sqrt{L}}\,\phi - \sqrt{L}\,\lambda_3 \, \phi \, H^{\dagger}\,H \,\delta (u)
   - L\,\la_4 \, \phi^2 \, H^{\dagger}\,H\,\delta (u)\,,
\eea
where $A,B=\{\mu,5\}$ and $G^{AB}$ is the 5D inverse metric tensor, with entries
\bea
G^{00}= h(u)~~;~~ G^{05}=G^{50}=g(u)~~;~~ G^{55}= G^{ii}=-1\,,
\eea
and all remaining entries zero.
From the 5D kinetic term, one sees that the mass dimension of $\phi$ is 3/2.
Constant factors of $\sqrt{L}$ have been inserted for later convenience,
so that the mass dimensions of $\la_1,\ \la_3$ and $\la_4$ have the usual 4D values of
3, 1, and 0, respectively.  The appearance of the delta function $\delta (u)$ in ${\cal L}^{(5D)}$
restricts the interactions with SM particles to the brane ($u=0$), to which the SM particles
(here, the SM Higgs doublet $H$) are confined.
A consequence of the restriction of SM particles to the brane via the delta function is that
translation invariance in the $u$-direction is broken.  This means that neither KK number
nor particle momentum in the $u$-direction are conserved; overall momentum conservation is
restored when the recoil momentum of the brane is included.

\subsubsection{Non-renormalizable, Effective Field Theory}
\label{subsubsec:effective}
The model in Eq.~\rf{L5D_I} is power-counting renormalizable.
However, the broken translational invariance in the extra dimension(s) leaves the model non-renormalizable.
For example, the operators $\phi^4$ (of dimension six and so manifestly not renormalizable in 5D)
and $\phi^3$ (5D-renormalizable, but destabilizing the Hamiltonian until large $\phi$ values allow the $\phi^4$ operator to dominate)
are each induced on the brane by a virtual loop of $H$-field,
and they are increasingly divergent as the number of extra dimensions increases.
%
%
The fact that $\phi^4$ is a non-renormalizable operator, yet necessarily induced by the
operator $\phi^2 \,H^\dag\, H \,\delta(u)$ which is power-counting renormalizable,
is an indication that ${\cal L}^{(5D)}$ describes an effective theory on the brane, not a renormalizable theory.

The induced operators $\phi^4$ and $\phi^3$ on the brane do not affect the physics of interest
in this paper, and so we do not consider these operators any further.
However, there are further effects of the effective theory that cannot be ignored.
For example, higher-order Higgs-pair operators $(H^\dag H)^N$ are induced
by a virtual loop of $\phi$-field.
The $N=1$ and $N=2$ operators may be renormalized by the SM counter-terms,
but higher-order operators introduce divergences for which there are no counter-terms.
If the model were renormalizable, these higher-order operators would be finite and calculable.
Instead, they are divergent, as we briefly illustrate in Appendix~\rf{app:efftheory}.
Consequently, the model is an effective theory, valid up to an energy cutoff of characteristic scale
$\Lambda\sim 1/ (\lambda_4\, L)$.

Interestingly, the complications on the brane do not pervade the bulk where the $H$-field vanishes.
Since $H=0$ in the bulk, there are no $H$-loops, and so no new induced operators.
In the bulk, $\phi$ is described by free field theory.

We note that since $\phi$ is a gauge-singlet, its mass $m$ is unrelated to spontaneous symmetry-breaking and is best
thought of as a free parameter. We further note that the Higgs singlet is largely unconstrained by known physics.
For example, gauge-singlets do not contribute to the $\rho\equiv (M_W/M_Z\,\cos\theta_w)^2$ parameter.

\subsubsection{Scales of Validity}
\label{subsubsec:scales}
In order construct a wave packet smaller than the size $L$ of the extra dimension,
we require $L\gg 1/\sqrt{s_{\rm LHC}}$.
Combined with the fact that our effective theory is valid only up to the cutoff $\Lambda\sim 1/(\lambda_4 \,L)$,
which we want to lie above $\sqrt{s_{\rm LHC}}$,
we arrive at a small bounded value for $\lambda_4$:
\beq{lambda4bnd}
\lambda_4 < \frac{1}{L\,\sqrt{s_{\rm LHC}}} \sim 10^{-2}\,\left( \frac{10^{-18}{\rm m}}{L} \right)\left(\frac{10 \,\textrm{TeV}}{\sqrt{s_{\rm LHC}}}\right) \ll 1\,.
\eeq

For the LHC energy scale to probe the extra dimension, we must assume that the size
of the extra dimensions $L$ is $\agt 1/\sqrt{s_{\rm LHC}}$.
Since the LHC is designed to probe electroweak symmetry, one may equivalently write
$L\agt 1/{\rm TeV}\agt 10^{-19}$~m for the LHC reach.
It is useful at this point to briefly review the bounds on the size of extra dimensions.
The strongest constraint on the ADD scenario comes from limits on excess cooling of supernova due to
KK graviton emission~\cite{Cullen} (analogous to limits from cooling by axion emission).
One extra dimension is ruled out.
For two extra dimensions, the lower bound on the fundamental Planck scale is 10~TeV
and the upper bound on the size of the extra dimensions is $L \lesssim 10^{-7}$ m
if the two extra dimensions are of the same size, easily within the reach
$L\agt 10^{-19}$~m at the LHC.
Consistency with the solar system tests of Newtonian gravity also requires
at least two extra dimensions~\cite{Kribs}.
While we have shown that a single extra-dimension is sufficient to admit
our class of CTCs, our construction does not disallow further extra dimensions.

\subsection{Klein-Gordon Solution and Energy--Momentum Dispersion Relation}
\label{subsec:DispersionRelation}
To develop the field theory of the Higgs singlet,
we will need the energy dispersion relation for the $\phi$ particle modes.
The dispersion relation can easily be obtained from the equation of motion for the
free $\phi$ field:
\beq{KGeqn}
G^{AB}\,\partial_{A}\,\partial_B\, \phi + m^2 \,\phi =0\,.\quad({\rm 5D\ Klein-Gordon\ equation})
\eeq
In fact, an inspection of Eq. \eqref{metric2} (and the definition of \,$\tilde{t}$\, in the footnote Eq.~\rf{tbar})
suggests that the general solution to this
5D Klein-Gordon (KG) equation for the $n^{th}$~energy-eigenfunction should take the form
\bea
\phi_n^{\rm (KG)}
&=& e^{-i \,E_n \,[\,t+\int_{0}^{u}\; g(u)\; du\,]} \; e^{i \,\vec{p}\cdot \vec{x}} \; e^{i\, \xi\, u }\,,
\eea
where $E_n$ is the energy of the $n^{\textrm{th}}$ mode (at fixed $\vec{p}$) and
$\vec{p}$ is the standard three-momentum along the
brane direction.
Since the extra dimension is compactified, we require $\phi_n (u+L) =\phi_n (u)$ which,
in turn requires that
\beq{xi}
\xi = \bar{g} \,E_n-\frac{2\pi\,n}{L}~~~~~~\textrm{with}~~~~ n=0,\,\pm 1,\,\pm 2,\, \ldots
\eeq
where the mean value $\bar{g}$ is defined in Eq.~\rf{gbar}.\footnote
{
From Eq.~\rf{genmetric} we also get mean
$\bar{h}=1-[ \bar{g}^2+\half\sum_n a_n^2 + \half \sum_n b_n^2 ]$.
We will not need this relation in the present paper.
}
Thus, the solution to the KG equation is given by
\bea
\label{KG}
\phi_n^{\rm (KG)} &=& e^{-i\, E_n \,t}\; e^{i \,\vec{p}\cdot \vec{x}} \;e^{-i \,E_n \,\int_{0}^{u}\;(g-\bar{g}) du}\;
e^{-i\, n\, u /R}\,,
\eea
where we have defined an extra-dimensional ``radius'' $R\equiv L/2\pi$ to streamline some notation.

To determine the energy dispersion relation, we simply need to plug Eq. \eqref{KG} into the 5D~KG
equation above and solve for $E_n$. A bit of algebra yields the quadratic dispersion relation
\beq{qdispersion}
(\bar{g}^2-1)\,E_n^2-2\,\bar{g}\,E_n\,\frac{n}{R}
          +\vec{p}^{\ 2}+\frac{n^2}{R^2}+m^2 = 0\,.
\eeq
Solving for $E_n$ then gives
\footnote
{We are grateful to A\@. Tolley for correcting an error in an earlier version of our KG equation,
and providing the dispersion relation which solves the corrected equation.
}
\footnote
{Note that $\bar{g}$ but not $\bar{h}=1-\overline{g^2}$ appears in the KG solution and in the dispersion relation.
This is related to the fact that a coordinate change may bring the metric to Minkowski form with no vestigial mention of $h$
but with a pathologic ``time'' containing a boundary condition depending on $\bar{g}$.
See Eqs.~\rf{diffs} and \rf{metric2}.
}
\bea
\label{E_n}
E_n=\frac{\bar{g}\,\frac{n}{R} \pm \sqrt{\frac{n^2}{R^2} -(\bar{g}^2-1)\,(\vec{p}^{\ 2}+m^2)}}{\bar{g}^2-1}\,.
\eea
Eq.~\rf{E_n} makes it clear that the mode energy $E_n$ depends on $\vec{p}$ as well as on $n$;
nevertheless, for brevity of notation, we will continue to use the ``fixed $\vec{p}$'' notation
for both $E_n$ and $\phi_n$.
 In order to ensure that $E_n$ is real, the condition $\frac{n^2}{R^2} >
(\bar{g}^2-1)\,(\vec{p}^{\ 2}+m^2)$ needs to be satisfied. Also, we discard the case with $\bar{g} < 0$ which leads to
negative $E_n$. Both the reality and positive definiteness of $E_n$ are required to provide the stable modes
for the CTC geodesics.

From Eq.~(\ref{qdispersion}), we may also derive a lower bound on the time-traveling particle's energy.
The result is
\beq{Enbnd}
E_n > \lim_{\bar{g} \to 1} E_n = \frac{1}{2} \left( \frac{n}{R}+\frac{\vec{p}\,^2+m^2}{(n/R)} \right) \,.
\eeq

The dispersion relation for $E_n$ is interesting in several respects.
First of all, due to the time-independence of the metric $G_{AB}$ and the time-translational invariance
of the Lagrangian $\mathcal{L}^{(5D)}$, the energy $E_n$ of the particle is constant during its propagation
over the extra dimensional path (the bulk) which forms the CTC. In other words, the energy is covariantly conserved.
Secondly, it is only for the zero-mode ( with $n=0$ and $\bar{g}$ effectively zero in Eq.~\rf{E_n} )
that the dispersion relation is trivial.
The KK modes ($n\ne 0$) exhibit a contribution $n^2/R^2$ to the effective 4D mass-squared,
a complicated dependence on $\bar{g}$,
and a resultant ``energy offset'' $ (\frac{\bar{g}}{1-\bar{g}^2})\,\frac{n}{R}$
which arises from the off-diagonal, non-static nature of the metric.

Not surprisingly, the integer mode number $n$ has a quantum interpretation.
It is the number of full cycles of $\Phi_n^{\rm (KG)}$ commensurate with the
circumference $L=2\pi\,R$ of the extra dimension.
We see this in the following way:
The half-cycles of $\Phi_n^{\rm(KG)}$ are separated from $u=0$ by the distance $u_k$,
where $u_k$ is the solution to
\beq{uk-equation}
E_n\int_0^{u_k} du\,(g-\bar{g}) + \frac{2\pi\,n\,u_k}{L} = k\pi\,,
\quad k=1,\ 2,\ \dots
\eeq
Notice that the lengths of these half-cycles are not uniform.
However, the total number of half-cycles is obtained by setting
$(u_k)_{\max} = L$, for which Eq.~\rf{uk-equation} becomes simply
$2\pi\,n = k\pi$.
Thus, $k_{\max} = 2\,n$, and the number of full cycles is $k_{\max}/2$, which is $n$,
identical to the number of wavelengths commensurate with $L$ in the usual flat space
($g(u)\equiv 0$) case.  We conclude that a non-zero $g(u)$ alters the lengths of the
cycles in the extra dimension, but does not alter their total number, which is $n$ for the mode $\Phi_n^{\rm (KG)}$.

The sum on mode number plays the same role in the extra dimension that $\int d^3\,\vec{p}$
plays in 4D.  Consequently,
%
%
an arbitrary field in the 5D spacetime can be expanded as a linear combination of mode fields:
\beq{arbfield}
\phi(x^\mu, u)= \sum_{n=-\infty}^{\infty}\,\int d^3\,\vec{p}~~
  \tilde{\phi}_n (E_n, \vec{p}) \,\,\phi_n^{\rm (KG)}\,,
\eeq
where $\tilde{\phi}_n (E_n, \vec{p})$ are the weight functions of $n$ and $\vec{p}$.
This completes the construction of the scalar field in 5D
with a periodic boundary condition in the $5^{th}$ dimension.

\subsection{Wave Packet as a Sum of Many Modes}
\label{subsec:wavepacket}
The minimum quantum energy of the $n^{th}$ mode is
associated with motion purely in the compactified dimension.
Thus we take the $\vec{p}\rarr 0$ limit of the dispersion relation to determine
this minimum quantized energy.
From Eq.~\rf{E_n}, we have
\bea
\label{Eu}
E_n(\vec{p}=0) &=&
\left( \frac{n}{R} \right)
\left( \frac {\bar{g}+\sqrt{1+(1-\bar{g}^2)\,\frac{m^2 R^2}{n^2}} }{ 1-\bar{g}^2} \right) \\
  && \nonumber \\
  & \stackrel{z\,\equiv\, m^2 R^2/n^2\ll 1}{\longrightarrow} &
    \left( \frac{n}{R}\right) \left(\frac{1}{1-\bar{g}}+\frac{z}{2}\right) +{\cal O}\left(z^2\right)\,.\nonumber
\eea
So for $m^2\ll n^2/R^2$, we find an energy spectrum rising (nearly) linearly in $n$.
This means that the first $n_{\max}\sim \sqrt{s_{\rm LHC}}\,R\,(1-\bar{g})$ modes are excitable,
in principle.
In practice, decay of the SM Higgs to $\phi\,\phi$, or mixing of $H$ with $\phi$,
will excite many KK modes of $\phi$.
We have $n_{\max}\sim M_h\, R \sim 6\times 10^{10}\,(R/10^{-7}{\rm m})\,(M_h/125\ {\rm GeV})$.
Thus, we do not expect single or few mode excitations to be relevant.

If a single mode were excited, its wave function would span the entire
compactified interval $[0,\ L]$, analogous to a plane wave in 4D.
With a single mode, one would expect quantum mechanics rather than
classical concepts to apply.
However, when more modes are excited, which we expect to be relevant case,
their weighted sum may form a localized wave packet in  $[0,\ L]$,
in which case the deductions for a classical particle
in the earlier sections should apply.
We denote the relevant many-mode wave packet by $\phi_{\nbar}$,
where $\nbar$ is meant to be a typical or mean mode number of the packet.

However, even an initially localized wave packet will spread in time.
Such packet spreading does no harm to our conclusion --
that the secondary vertex of the co-rotating Higgs singlet will still precede the production vertex in time.
The spreading of the wave function just increases the variance of the distribution in negative $t$.
The classical equation of motion for the $u$-direction continues to describe the group velocity of the
centroid of the wave function as it travels in the $u$-direction.
The same happens for the tau and $b$~ fermions in Minkowski space,
as they progress from their production vertices to their decay vertices.

To understand the variance of the distribution of times between
primary and secondary vertices, we now quantify the wave function spreading.
To be explicit, we adopt a Gaussian wave-packet at t=0
with initial spatial spread $\Delta L_0$ in the 5th dimension.
The standard formula for wave packet spread in a single dimension is
$(\Delta L)^2 = (\Delta L_0)^2 + (\tau/m\,\Delta L_0)^2$.
Here $L$ is the circumference of the extra dimension as usual,
and $\tau$ is the proper time of the wave function ({\sl a priori} independent of the time of the observer).
There are two characteristic times of interest to us.
The first is the time at which the packet begins to noticeably spread, given by $\tau_1\equiv m\,(\Delta L_0)^2$.
The second is the time when the packet completely fills the compactified dimension, given implicitly by
$\Delta L(\tau_2)=L$, and explicitly by
$\tau_2 = m \,(\Delta L_0)\,L = \left(\frac{L}{\Delta L_0}\right) \tau_1$.
For an experimental energy such that the $n^{\rm th}$ mode is excitable, we have shown below Eq. \eqref{uk-equation} that there are $n$ full
cycles within the extra dimension.
Each mode is a distorted plane wave filling the compact dimension, with an initial width of roughly $L$.
Adding more modes decreases the width.
We approximate the initial width of the Gaussian wave-packet with $n$ modes to be roughly $\Delta L_0 = L/\nbar$,
where again, $\nbar$ is the mean mode number.
We further approximate $m \sim 2\,\pi\,\bar{n}/L$,
and arrive at $\tau_1\sim 2\,\pi\, L/\bar{n}$.
Finally, taking $\bar{n}\ll L\sqrt{s_{\rm LHC}}$,
say, $\sim (L\sqrt{s_{\rm LHC}}/10) \sim L\,({\rm TeV})\sim L/10^{-19}{\rm m}$,
we find $\tau_1\sim 100/\sqrt{s_{\rm LHC}}\sim 10^{-27}$~s and
$\tau_2\sim \bar{n}\,\tau_1 \sim  (L/10^{-19}{\rm m})\times 10^{-27}$~s.
Lab frame time  $t$ is related to time in the wave function frame by $t=\gamma_u\, \tau$,
with $\gamma_u$ being the Lorentz factor for a boost in the $u$-direction.
However, $\gamma_u$ is nowhere near large enough to compensate for the
many orders of magnitude needed to qualitatively change the results just obtained for wave packet spreading.
We conclude that the times $t_1$ and $t_2$ which characterize the wave function spreading in the lab frame
are much shorter than the $\agt$~picosecond time associated with displaced vertices.
Consequently, the wave packet effectively spreads linearly in time with coefficient $(m\,\Delta L_0)^{-1} \sim 1/2\pi$,
creating a considerable variance in the times
(negative for co-rotating Higgs singlets and positive for counter-rotating Higgs singlets)
between primary and secondary vertices.

We make here a side remark that in addition to the minimum energy
associated with motion in the $u$~direction,
the momentum in the $u$-direction is also interesting.
While not observable, it is of sufficient mathematical interest that we
devote Appendix~\rf{app:p5} to its description.

We have seen that the localized time-traveling particle is a sum over many modes.
The Lagrangian describing its production, which we now turn to, is also a sum over modes,
with each mode characterized by an energy $E_n$ according to our dispersion relation, Eq.~\rf{E_n}.
The weight functions in the Lagrangian are all unity.
That is to say, calculations begin in the usual fashion, as perturbations about a free field theory.

\subsection{4D Effective Lagrangian Density}

The reduction of the 5D theory to an effective 4D Lagrangian density is accomplished by the integration
\bea
\label{L4D}
\mathcal{L}^{(4D)}= \int_0^L\,du\; \left(\,\mathcal{L}_0 + \mathcal{L}_I\,\right)\,,
\eea
where $\mathcal{L}_0$ and  $\mathcal{L}_I$ are the 5D free and interacting Lagrangian densities
given in Eqs.~\rf{L5D_0} and \rf{L5D_I}.
We are interested in showing how the SM Higgs
interacts with the singlet Higgs' energy eigenstates $\phi_n (x^{\mu},u)$.
Thus, the explicit expression for the Lagrangian density of the free singlet $\int_0^L\,du\; \mathcal{L}_0$
is irrelevant for the following discussions.

Now we turn to the interaction terms.
Neglecting the tadpole term
$-\frac{\lambda_1}{\sqrt{L}}\,\phi$
(which can be renormalized away, if desired), we have for the 4D interaction Lagrangian density
\bea
\int_0^L\,du\; \mathcal{L}_I &=& -\sqrt{L}\,\lambda_3 \,  H^{\dagger}\,H \,\int_0^L\,du\; \delta (u)\;
\sum_{n}\, \phi_n(x^{\mu},u) \nonumber \\
&&  - L\,\la_4 \, H^{\dagger}\,H\,\int_0^L\,du\; \delta (u)\; \sum_{n_1,n_2}\,
\phi_{n_1}(x^{\mu},u) \, \phi_{n_2}(x^{\mu},u)
\nonumber \\
&=& -\lambda_3 \, H^{\dagger}\,H \, \sum_{n}\, \sqrt{L}\,\phi_n(x^{\mu},0) - \la_4 \, H^{\dagger}\,H\, \sum_{n_1,n_2}
\,\sqrt{L}\, \phi_{n_1}(x^{\mu},0)\,\sqrt{L}\,\phi_{n_2}(x^{\mu},0) \nonumber \\
\label{L4D1}
&=& -\lambda_3 \, H^{\dagger}\,H \, \sum_{n}\,\bar{\phi}_n - \la_4 \, H^{\dagger}\,H\, \sum_{n_1,n_2}
\, \bar{\phi}_{n_1}\,\bar{\phi}_{n_2}\,,
\eea
where $\bar{\phi}_n = \sqrt{L}\,\phi_n(x^{\mu},0)$ is the singlet field on the brane,
normalized with $\sqrt{L}$ to its 4D~canonical dimension of one.
Note that since the energy is covariantly conserved,
both $\phi_n(x^\mu,u)$ and $\phi_n(x^\mu, 0)$ will have the same energy $E_n$.

%
%

\subsection{Incorporating Electroweak Symmetry Breaking}
\label{subsec:EWSB}
Electroweak symmetry breaking (EWSB) in ${\cal L}^{(4D)}$ is effected by the replacement
$H^{\dagger} H \rarr \frac12 (h+v)^2$ in Eq.~\eqref{L4D1}, where $v\sim 246$~GeV is the SM Higgs vev.
The result is
\beq{L4D2}
\mathcal{L}^{(4D)} = \int_0^L\,du\; \mathcal{L}_0 - \frac{\lambda_3}{2} \,(\,2 \, v \, h + \,h^2\,)\,\sum_n \bar{\phi}_n
  - \frac{\la_4}{2}\,(v^2+2\,v\,h+h^2\,)\,\sum_{n_1,n_2} \bar{\phi}_{n_1}\bar{\phi}_{n_2}\,.
\eeq
Omitted from Eq.~\rf{L4D2} is a new tadpole term
$-\,\half\, \la_3 \,v^2\,\sum_n \,\bar{\phi}_n $ linear in \,$\bar{\phi}_n$.
It is irrelevant for the purposes of this article,
so we here assume for simplicity that it can be eliminated by fine-tuning the corresponding counter-terms\footnote
{A different theory emerges if the counter-term is chosen to allow a nonzero tadpole term.
For example, the singlet $\phi$ field may then acquire a vev.}.
The off-diagonal terms in $\la_4\,\sum_{n_1,n_2}v^2\,\bar{\phi}_{n_1}\,\bar{\phi}_{n_2}$ mix
different $\bar{\phi}_n$ fields, while the terms $\sum_{n}\la_3\,v\,h \,\bar{\phi}_n$ induce singlet-doublet mixing.
We make the simplifying assumption that $(\la_3\,v,\,\la_4 \, v^2) \ll m^2+\frac{n^2}{R^2}$,
so that upon diagonalization of the mass-matrix,
the mass-squared of the $n^{\rm th}$ KK mode remains close to $M_n^2 \approx  m^2+\left(\,\frac{n}{R}\,\right)^2$.
We emphasize that this assumption is made so that the calculation may proceed to a more complete proof of principle
for acausal signals at the LHC.
In fact, it seems likely to us that acasual signals are inherent in the present model
even without this simplifying assumption, and probably in other models not yet explored.

We now turn to the details of Higgs singlet production and detection at the LHC.
As encapsulated in Eq.~\rf{L4D2},
Higgs singlets can be produced either from decay of the SM Higgs
or through mass-mixing with the SM Higgs.
We discuss each possibility in turn.

\section{Phenomenology of Pre-Appearing Secondary Vertices}
\label{sec:pheno}
%
Motivated by the advent of the LHC, we will next discuss the production and detection at the LHC of
Higgs singlets which traverse through the extra dimension and violate causality.

How would one know that the Higgs singlets are crossing and re-crossing our brane?
The secondary vertex may arise from scattering of the singlet, or from decay (if allowed by symmetry)
of the singlet.
These ``vertices from the future'' would appear to occur at random times, uncorrelated with the pulse times of
the accelerator.\footnote
{
Pre-appearing events might well be discarded as ``noise''.
We want to caution against this expediency.
}
The essential correlation is via momentum.
Exactly the three-momentum missing from the primary vertex is restored in the secondary vertex.
Of course, the singlet particles on counter-rotating geodesics will arrive
back at our brane at later times rather than earlier times.
The secondary vertices of counter-rotating particles will appear later than the primary vertices
which produced them, comprising a standard ``displaced vertex'' event.

The rate of, distance to, and negative time stamp for, the secondary vertices will depend on
three parameters.
First is the production rate of the Higgs doublets, which is not addressed in this paper.
Secondly is the probability for production of the Higgs singlet per production of the Higgs doublet,
which we denote as $\PP$.  Thirdly is the probability for the Higgs singlet to interact, either by
scattering or by decaying, to yield an observable secondary vertex in a detector.
Of course, for the Higgs singlet to scatter on or decay to SM particles (via coupling with the SM Higgs doublet),
the singlet must be on the brane.
We define  $\PD$ to be the probability for the Higgs singlet to create a secondary vertex per brane crossing.

Since $\phi$ is a singlet under all SM groups, it will travel almost inertly through the LHC detectors.
Each produced singlet wave-packet $\bar{\phi}_{\nbar}$ exits the brane and propagates into the bulk,
traverses the geodesic CTCs, and returns to cross the brane at times $t_N$ given by Eq.~\eqref{tNagain}.
Classically, translational invariance in the $u$-direction is broken by the existence of our brane,
and so $u$-direction momentum may appear non-conserved.\footnote
{Momentum in fact is conserved in the following sense:
From the 5D point of view, energy-momentum is conserved as the brane recoils against the emitted Higgs singlet.
From the 4D point of view, energy-momentum is conserved when the dispersion relation of Eq.~\rf{E_n} is
introduced into the 4D phase-space, as is done in Appendix~\rf{app:SMHiggsDecay}.
}
The classical picture that emerges is restoration of $u$-momentum conservation when brane recoil is included.

It is worth noting that all equations from the first six sections of this paper are classical equations,
and so are independent of mode number $n$.  Thus, these equations apply to the complete wave packet
$\phi_{\nbar}$ formed from superposing many individual modes.

The probability for the Higgs singlet, once produced with probability $\PP$,
to survive $N$~traversals of the extra-dimension and ``then'' decay or scatter on the $(N+1)^{th}$ traversal is
\beq{dkNtraversals}
P(N+1)=\PD\,(1-\PD)^N\sim \PD\,e^{-N\PD}\,.
\eeq
The latter expression, of Poisson form, pertains for $\PD\ll 1$, as here.
It is seen that even small scattering or decay probabilities per crossing exponentiate over
many, many crossings to become significant.
For this Poissonian probability, we have some standard results:
the probability for interaction after $N$ traversals is flat up to the mean value
$\langle N\rangle=1/\PD$ (very large),
the rms deviation, $\sqrt{\langle N^2\rangle-\langle N\rangle^2}$, is again $1/\PD$
(very wide, as befits a flat distribution),
and the probability for the singlet to interact in fewer crossings than
$\langle N\rangle=1/\PD$ is $1-e^{-1}=63\%$.

Thus, the typical negative time between the occurrence of the primary vertex and the pre-appearance of the
secondary vertex should be, from Eq.~\rf{tNagain}, of order
\beq{pre-time}
|t_{\langle N\rangle}|=\langle N\rangle\,|t_1| \,\sim\,\frac{L/c}{\PD}\,.
%
\eeq
The typical range of the secondary vertex relative to the production site, from Eq.~\rf{branecoord2} is
\beq{secrange}
r_{\langle N\rangle}=\frac{\cot\theta_0\,L}{\PD}\,.
\eeq
(Recall that $\theta_0$ is the exit angle of the singlet relative to the brane direction.)
The probability (per unit SM Higgs production)
for the Higgs singlet to be produced and also interact within a distance $l$ of the production site
is then
\begin{eqnarray}\label{secondary<l}
P(\textrm{range of secondary vertex}<l) &\approx&\PP\,\int_0^{\frac{l}{\cot\theta_0\,L}\,\PD} d (N\PD)\,e^{-(\,N\PD\,)} \nonumber \\
    &=& \PP\,\left[\,1-e^{(l\,\tan\theta_0 /L\,)\,\PD}\,\right]\,,
\end{eqnarray}
which provides the limiting value
\beq{limitsecondary}
P(\textrm{range of secondary vertex}<l)\approx \PP\,\PD\frac{l\,\tan\theta_0}{L}\,,
   \hspace{1.0cm}{\rm for\ \ } \frac{l\,\tan\theta_0}{L}\,\PD\ll 1\,.
\eeq
For the secondary vertex to occur within the LHC detectors, one requires $l=l_{\rm LHC}\sim 10$~m.

We will assume that $\langle\tan\theta_0\rangle$ is of order unity.  Then the
figure of merit that emerges for CTC detection is $\PP\,\PD\,l_{\rm LHC}/L$.
We have seen that the maximum allowed value of $L$ for two extra dimensions is $10^{-7}$~m,
and that the reach of the LHC is $\sim 1/\sqrt{s}\sim 10^{-19}$~m.
Thus, we are interested in an extra-dimensional size $L$ within the bounds $[ 10^{-19},\,10^{-7} ]$~m.
Below we shall see that the acausal pre-appearance of the secondary vertex
for the co-rotating singlet may be observable at the LHC.

The Higgs singlet production and interaction mechanisms
depend on the symmetry of the Higgs singlet-doublet interaction terms in the Lagrangian.
Therefore the production and detection probabilities $\PP$ and $\PD$, respectively, do as well.
We discuss them next.  There are two possibilities for our 5D Lagrangian, with and without
a $Z_2$ symmetry $\phi\leftrightarrow -\phi$.

\subsection{Without the {\boldmath $\phi \,\leftrightarrow\, -\phi$} Symmetry}
\label{subsec:withoutZ2}

In this subsection, we ignore the possible $Z_2$ symmetry and keep the trilinear term
$\la_3\, \, H^{\dagger}\,H \,\sum_{n}\,\bar{\phi}_n$ in the Lagrangian density.
When the SM Higgs acquires its vev $v$, we have the resultant singlet-doublet mass-mixing term
$\la_3\,v\,h\,\sum_{n}\,\bar{\phi}_n$ in the 4D Lagrangian of Eq.~\rf{L4D2}.
Note that this singlet-doublet mixing can only occur when the singlet particle \,$\bar{\phi}_n$\, is traversing the brane,
as the field $H$ is confined to the brane.

In Refs.~\cite{Giudice,Gunion}, it was shown that mixing of the Higgs field with higher-dimensional graviscalars
enhances the Higgs invisible width $\Gamma_{\rm inv}$ while maintaining the usual Breit-Wigner form.
The invisible width is extracted from the imaginary part of the Higgs self-energy graphs, which
includes the mixing of the Higgs with the many modes.
These calculations apply in an analogous way to the Higgs--many-mode mixing in our model.
Thus the many-mode wave function $\bar{\phi}_{\nbar}$, which we introduced in section~\rf{subsec:wavepacket},
is the Fourier transform of an energy-space Breit-Wigner form.
In practice, this means that the sum includes all modes within the Higgs invisible width,
sculpted by the Breit-Wigner shape.
Including modes with energy from $M_h-\Gamma_{\rm inv}/2$ to $M_h+\Gamma_{\rm inv}/2$,
we have a mean mode number $\nbar \sim  M_h\, R$, and an effective coupling for $h$-$\bar{\phi}_{\nbar}$ mixing of
$-m_{\rm mix}^2\sim -\lambda_3 \,v\, R\,\,\Gamma_{\rm inv}$,
since $1/R$ is roughly the energy spacing between modes.
In the model of~\cite{Giudice}, the branching ratio
$BR(h\rarr {\rm invisible})$
is calculated and shown to vary from nearly one with two extra dimensions,
to three orders of magnitude less with six extra dimensions.
We may expect something similar here.

Diagonalization of the effective $2\times 2$ mass-mixing matrix leads to the
mixing angle \,$\theta_{h\bar{\phi}_{\nbar}}$\, between $h$ and $\bar{\phi}_{\nbar}$.
We assume this angle to be small, an assumption equivalent to assuming $m_{\rm mix}^2 \ll M^2_h$.
%
%
We label the resulting  mass eigenstates of this $2\times 2$ subspace \,$h_2$\, and \,$h_1$\, with masses $M_2$ and $M_1$.
The mass eigenstates are related to the unmixed singlet and doublet states \,$\bar{\phi}_{\nbar}$\, and \,$h$\, by
\be
\left(
\begin{array}{c}
|h_2\rangle \\
|h_1\rangle
\end{array}
\right) =
\left(
\begin{array}{cc}
\cos\thet & \sin\thet \\
-\sin\thet & \cos\thet
\end{array}
\right)
\left(
\begin{array}{c}
|h\rangle \\
|\bar{\phi}_{\nbar}\rangle
\end{array}
\right)\,,
\ee
and the inverse transformation is
\be
\left(
\begin{array}{c}
|h\rangle \\
|\bar{\phi}_{\nbar}\rangle
\end{array}
\right) =
\left(
\begin{array}{cc}
\cos\thet & -\sin\thet \\
\sin\thet & \cos\thet
\end{array}
\right)
\left(
\begin{array}{c}
|h_2\rangle \\
|h_1\rangle
\end{array}
\right)\,.
\ee
We will assume for definiteness that on the brane, the two states (in either basis)
quickly decohere due to a significant mass splitting.
This assumption is reasonable since the decoherence time is
$t_{\rm deco}\sim \frac{2\pi\gamma_u}{\delta M}\sim10^{-23}\,\left({\rm GeV}/\delta M\right)$~s.
(The differing mass peaks $M_2$ and $M_1$ may thus be distinguishable at the LHC.)
So we consider only the classical probabilities \,$\cos^2\thet$\, and \,$\sin^2\thet$
in the remaining calculation.

The electroweak interaction, which would otherwise produce the SM Higgs,
will now produce both mass eigenstates \,$h_2$\, and \,$h_1$\,
in the ratios of \,$\cos^2\thet$\, and \,$\sin^2\thet$, times phase space factors.
For purposes of illustration, we take these phase space factors to be the same for both modes.
The \,$\bar{\phi}_{\nbar}$\, components of these mass eigenstates \,$h_2$\, and \,$h_1$
are given by the probabilities \,$\sin^2\thet$\, and \,$\cos^2\thet$,\, respectively.
Thus, per production of a Higgs doublet,
the probability that a singlet $\bar{\phi}_{\nbar}$ is produced is
$\PP= 2\sin^2\thet\,\cos^2\thet=\half\sin^2(2\,\thet)\stackrel{\thet\ll 1}{\longrightarrow}\sim 2\,\thet^2$.

%
%

Upon returning to the brane, these pure \,$\bar{\phi}_{\nbar}$\, states mix again and hence split
into \,$h_2$\, and \,$h_1$\, states,
with respective probabilities \,$\sin^2\thet$\, and \,$\cos^2\thet$.\,
The probabilities for these \,$h_2$\, and \,$h_1$\, states
to decay or interact as a SM Higgs \,$h$\, are respectively given by \,$\cos^2\thet$\, and \,$\sin^2\thet$.\,
Thus, the total probability per returning \,$\bar{\phi}_{\nbar}$\, particle
per brane crossing to decay or interact
as a SM Higgs \,$h$\, is again $\PD= \PP= \half\sin^2(2\,\thet)\sim 2\,\thet^2$, a very small number.
%
%
Therefore, per initial Higgs doublet production
the probability for a singlet \,$\bar{\phi}_{\nbar}$\, component
to be produced and to acausally interact on the $N^{th}$ brane-crossing is approximately \,$4\,\thet^4$,\,
nearly independent of the number of brane-crossings.
These brane-crossings happen again and again until the interaction ends the odyssey.
From the initial production of the \,$\bar{\phi}_{\nbar}$\, component to its final interaction upon brane-crossing, the time elapsed (as measured
by an observer on the brane) is again given by $t_N$ in Eq. \eqref{tNagain}.


Therefore, in the broken $\phi\leftrightarrow -\phi$ symmetry model,
we expect the probability that a pre-appearing secondary
vertex will accompany each SM Higgs event to be
\begin{eqnarray}
\label{P_range}
P(\textrm{range of secondary vertex}<l_{\rm LHC}) &\approx& \PP\,\PD\,(10\ {\rm m})/L \nonumber \\
&\sim& 10^8\,(2\,\thet^2)^2\,\left(\frac{10^{-7}{\rm m}}{L}\right).
\end{eqnarray}
Here, the negative time between the secondary and primary vertices would be
\begin{equation}
\label{t_negative}
|t_{\langle N\rangle}|\approx \frac{L}{\PD}\sim 3\times 10^{-16}\,\left(\frac{L / 10^{-7}{\rm m}}{2\,\thet^2}\right)~s.
\end{equation}

Observability of a negative-time secondary vertex requires that
$|t_{\langle N\rangle}|$ lies in the interval of roughly a picosecond to 30~nanoseconds,
and that the probability per Higgs doublet $P(\textrm{range of secondary vertex}<l_{\rm LHC})$ exceeds roughly one per million.
Manipulation of Eqs. \eqref{P_range} and \eqref{t_negative} then reveals that the two observability requirements are met
with any $L$ down to $10^{-14}$~m (as discussed
in Section~\rf{subsec:DispersionRelation}, we require $L < 10^{-7}$~m to avoid excessive supernova cooling),
and
\bea
10^{-7}\,\sqrt{\frac{L}{10^{-7}\,\textrm{m}}} \le 2\,\thet^2 \le
\left\{
\begin{array}{ll}
   10^{-3.5}\,\left(\,\frac{L}{10^{-7}\,\textrm{m}}\,\right), & \hbox{ \textrm{if}  $L > 10^{-8}\,\textrm{m}$;} \\
   10^{-4} \,\sqrt{\frac{L}{10^{-7} \,\textrm{m}}}, & \hbox{~\textrm{if}  $L < 10^{-8}\, \textrm{m}$.}
  \end{array}
\right.
\eea

For example, with the largest value of $L$ allowed by SN cooling rates for two extra dimensions,
$10^{-7}$~m, one gets
$10^{-7}\le 2\,\thet^2 \le 10^{-3.5}$.
With the smallest value of $L$ allowed for observability in the LHC detectors,
$10^{-14}$~m, one gets
$10^{-10.5}\le 2\,\thet^2 \le 10^{-7.5}$.

Thus, we have demonstrated that for a range of choices for $L$ and $\thet$, or equivalently,
for $\PP$ and $\PD$, pre-appearing secondary vertices are observable in the LHC detectors.


\subsection{With the {\boldmath $Z_2$} Symmetry {\boldmath $\phi \,\leftrightarrow\, -\phi$}}
\label{subsec:withZ2}

If one imposes the discrete $Z_2$ symmetry $\phi \,\leftrightarrow\, -\phi$,
then the coupling constants \,$\la_1$ \,and\, $\la_3$\, are zero\footnote
{
The 4-dimensional counterpart of this simple $Z_2$ model was first proposed in~\cite{Zee}, where the
$\phi$ quanta are called ``scalar phantoms''.
}
and the low mode Higgs singlets are stable, natural, minimal candidates for weakly interacting massive particle
(WIMP) dark matter~\cite{Burgess,Bento}.
Constraints on this model from the CDMS II experiment \cite{CDMS2} have been studied in \cite{Asano,Farina}.
The discrete symmetry $\phi \,\leftrightarrow\, -\phi$ also forbids the Higgs singlet to acquire a vacuum expectation
value (vev). This precludes any mixing of the Higgs singlet with the SM Higgs.
With the $Z_2$ symmetry imposed, SM Higgs decay is the sole production mechanism of the Higgs singlet.
The decay vertex of the SM Higgs provides the primary vertex for the production of the Higgs singlet,
and subsequent scattering of the singlet via $t$-channel exchange of a SM Higgs provides the secondary vertex.

In Eq.~\rf{L4D2}, each term of the form $ \la_4\,v\,h\,\bar{\phi}_{n_1}\,\bar{\phi}_{n_2}$
provides a decay channel for the SM Higgs into a pair of Higgs singlet modes,
if kinematically allowed.
The general case $h\rightarrow \bar{\phi}_{n_1} \,\bar{\phi}_{n_2}$ with $n_1\neq n_2$ is considered
in Appendix~\rf{app:SMHiggsDecay}.
Here we exhibit the simplest decay channels to single mode states,
$h\rightarrow \bar{\phi}_{n} \,\bar{\phi}_{-n}$
and $h\rightarrow \bar{\phi}_{n}\,\bar{\phi}_{n}$.
The width for $h\rightarrow \bar{\phi}_{n} \,\bar{\phi}_{-n}$ is
\bea
\label{decayrate_n-n}
\Gamma_{h \rightarrow \bar{\phi}_n\, \bar{\phi}_{-n}} = \frac{\la_4^2\,v^2}{16 \,\pi\,M_h }\,\beta_{n,-n}~~;
~~~~\beta_{n,-n}=\frac{1}{\sqrt{1-\bar{g}^2}}\,\sqrt{1-\frac{4\,\bar{M}_n^2}{(1-\bar{g}^2)^2\,M_h^2}}\;,
\eea
while the width for $h\rightarrow \bar{\phi}_{n} \,\bar{\phi}_{n}$ is
\bea
\label{decayrate_n+n}
\Gamma_{h \rightarrow \bar{\phi}_n\, \bar{\phi}_{n}} = \frac{\la_4^2\,v^2}{8 \,\pi\,M_h }\,\beta_{n,n}~~;
~~~~\beta_{n,n}=\frac{1}{\sqrt{1-\bar{g}^2}}\,\sqrt{1-\frac{4\,\bar{M}_n^2}{\left(\,(1-\bar{g}^2)\,M_h-\frac{2\,\bar{g}\,n}{R}\,\right)^2}}\;,
\eea
where $\bar{M}^2_n= (1-\bar{g}^2)\,m^2 +\frac{n^2}{R^2}$.

%
%

The above formulae apply to single mode final states.
Ref~\cite{Gunion} looked at Higgs decay to a pair of graviscalars.
The authors found via a quite complicated calculation that the decay was suppressed compared to simpler Higgs-graviscalar mixing.
However, their model concerned gravitational coupling, whereas our model has completely different
couplings for mixing. 
Thus, the techniques of~\cite{Gunion} may apply, but the conclusions do not.
We choose to finesse the hard calculation with an order of magnitude estimate.
Each sum on modes is constrained by phase space
(and not by $\Gamma_{\rm inv}$ as in the broken-$Z_2$ mixing case),
and so includes roughly $n_{\max}\sim M_h\,R/2 \sim 3 \times 10^{10} \,(R/10^{-7}{\rm m})\,(M_h/125\, {\rm GeV})$ states.
A typical mode value will be $\nbar\sim  M_h \,R/4$.
Thus, from here forward, in Eqs.~\rf{decayrate_n-n} and \rf{decayrate_n+n},
we set $n$ to $\nbar$ taken as $M_h\, R/4$,
and multiply the RHS by the mode-counting factor $n_{\max}$ for each of the final state singlets,
yielding the rate-enhancing factor $n_{\max}^2 \sim  2.3 \times 10^{19}\,\left(L/10^{-7}{\rm m}\right)^2 \,\left(M_h/125\,{\rm GeV}\right)^2$.

It is illuminating to look at the ratio of decay widths to $\bar{\phi}_{\nbar}$ pairs and to $\tau$-lepton pairs.
For the $\tau$, the coupling $g_Y$ to the SM Higgs is related to the $\tau$ mass through EWSB:
$g_Y^2=2\:m_\tau^2/v^2$.
Neglecting terms of order $(m_\tau/M_h)^2$, the ratio can be approximated as
\bea
\frac{\Gamma_{h \rightarrow \bar{\phi}_{\nbar}\,\bar{\phi}_{\pm\nbar}}}{\Gamma_{h \rightarrow \tau^+\tau^-}} \sim
\frac{\la_4^2\, v^4\,n_{\max}^2}{M_h^2\,m_\tau^2}\,\beta_{\nbar,\pm \nbar} \,.
\eea
This ratio can be much greater than unity, even for perturbatively small $\la_4$, and so $\PP$
can be nearly as large as unity.  It thus appears likely that $\bar{\phi}_{\nbar}$ particles
will be copiously produced by SM Higgs decay if kinematically allowed,
%
that their KK modes will explore extra dimensions if the latter exist,
and finally, that the KK modes will  traverse the geodesic CTCs,
if nature chooses an appropriately warped metric.

The exact $Z_2$ symmetry of the model under consideration forbids decay of the lighter $\phi$ singlets.
The $Z_2$-model does allow communication of the $\phi$ with SM matter through $t$-channel exchange of a SM Higgs.
The top-loop induced coupling of the SM Higgs to two gluons provides the dominant coupling of the SM Higgs $h$ to SM matter.
Despite the small couplings of $h$ to the SM, and $\lambda_4\,v$
at the $h\,\bar{\phi}_{\nbar_1}\,\bar{\phi}_{\nbar_2}$ vertex,
singlet scattering is enhanced by $n_{\max}$ in amplitude, and so $n_{\max}^2$ in rate.
Moreover, the singlet will eventually scatter since it will circulate through
the periodic fifth dimension again and again until its geodesic is altered by the scattering event.
The scattering cross section is of order
\beq{sigscatt}
\sigma_{\phi_{\nbar}\,\rm{N}}\sim\left\{\frac{(\lambda_4\,v\,n_{\max})\,
  \lambda(h\rarr t\bar{t}\rarr gg)\,(\alpha_s/4\pi)}{M_h^{2}}\right\}^2\,,
\eeq
where
$\lambda(h\rarr t\bar{t}\rarr gg)$ is the effective coupling of $h$ to the nucleon N through a virtual top-loop
at the Higgs end and two gluons at the nucleon end.
This coupling strength is of order $\frac{\alpha_s}{4\pi}\sim 10^{-2}$.
Thus, we expect
\beq{crossx}
\sigma_{\phi_{\nbar}\,\rm{N}}\sim \left[ \left(5\times 10^9\;\lambda_4\right)
  \left(\frac{125\,{\rm GeV}}{M_h}\right)\, \left(\frac{L}{10^{-7}{\rm m}}\right)\right]^2\,fb\,.
\eeq
We get the scattering probability per brane crossing by multiplying this cross section by
the physical length of the brane crossing $\sim w$, by the fraction of time spent on the brane $\Delta t/t\sim w/L$,
and by the target density $\rho$; the brane width $w$ is a free parameter, beyond our classical model,
but presumably of order $\sim L$.  We find
\beq{PDscattering}
\PD=3\times 10^{-20}\,\left(\frac{\sigma_{\phi_{\nbar} N}}{fb}\right)\left(\frac{\rho}{5\ {\rm g/cm}^3}\right)
    \left(\frac{w}{L}\right)^2\left(\frac{L}{10^{-7}{\rm m}}\right)\,.
\eeq
As a scaling law, we have $\PD\propto L\,w^2/M_h^2$, which grows linearly in $L$.

In summary, with the $\phi\leftrightarrow -\phi$ symmetry, we expect the probability that a pre-appearing secondary
vertex will accompany each Higgs event at the LHC to be
$\PP\,\PD\,(10\ {\rm m})/L \sim 10^{-12}
    \left(\frac{\sigma_{\phi_{\nbar} N}}{fb}\right)\left(\frac{\rho}{5\ {\rm g/cm}^3}\right)\left(\frac{w}{L}\right)^2$
for $\PP\sim 1$.
The negative time between the secondary and primary vertices would be
$\sim L/\PD\sim 10^4
   \left[\left(\frac{\sigma_{\phi_{\nbar} N}}{fb}\right)\left(\frac{\rho}{5\ {\rm g/cm}^3}\right)
   \left(\frac{w}{L}\right)^2\right]^{-1}$~s.
These numbers for the unbroken $\phi\leftrightarrow -\phi$ model are encouraging or discouraging,
depending on Nature's choice for the compactification length~$L$.
The model with broken $\phi\leftrightarrow -\phi$ symmetry is more encouraging.

%
%
%

\subsection{Correlation of Pre-Appearing Secondary and Post-Appearing Primary Vertices}
\label{subsec:vertices}
Finally, we summarize the correlations between the primary vertex producing the negative-time traveling
Higgs singlet and the secondary vertex where the Higgs singlet reveals itself. As we have seen above, the first
correlation is the small but possibly measurable negative time between the primary and secondary vertices.

The second correlation relating the pre-appearing secondary vertex and the post-appearing primary vertex
is the conserved momentum. As with familiar causal pairs of vertices, the total momentum is zero only
for the sum of momenta in both vertices.
Momentum conservation can be used to correlate the pre-appearing secondary vertex with its later
primary vertex, as opposed to the background of possible correlations of the secondary vertex with
earlier primary vertices.

Thus, the signature for the LHC is a secondary vertex pre-appearing in time relative to the associated
primary vertex.  The two vertices are correlated by total momentum conservation.
If such a signature is seen, then a very important discovery is made.
If such a signature is not seen, then the model is falsified for the energy scale of the LHC.

\section{Discussions and Further Speculations}

As we have just demonstrated with a simple model, it is possible to have a significant amount of KK Higgs singlets
produced by decay of, or mixing with, SM Higgs particle at the LHC.
If Nature chooses the appropriate extra-dimensional
metric, then these KK Higgs singlets can traverse the geodesic CTCs and thereby undergo travel
in negative time.\footnote
{
The idea of causality violation at the LHC is not new.
For example, a causality violating SM Higgs has been proposed in~\cite{Nielsen:2007ak},
by invoking an unconventional complex action.
The possibility of wormhole production at the LHC has been discussed in~\cite{Aref'eva:2007vk}.
The idea of testing the vertex displacements for the acausal Lee-Wick particles at the LHC has been proposed
by \cite{LeeWick}.
Also, some suggestive and qualitative effects associated with time traveling particles have been proposed
in~\cite{Mironov:2007bm}, but without any concrete LHC signatures.
}

One may wonder why such acausal particles, if they exist, have not been detected up to now.
One possible answer is that these time-traveling particles
may have been recorded, but either unnoticed or abandoned as experimental background.
Another possible answer could be that there has not been sufficient volume or instrumentation available to the detectors before
now to detect these events.
It may be that for the first time our scientific community has built accelerators
capable of producing time-traveling particles, and also detectors capable of sensing them.

One may also wonder whether an acausal theory could be compatible with quantum field theory (QFT).
After all, in the canonical picture, QFT is built upon time-ordered products of operators,
and the path integral picture is built upon a time-ordered path.
What does ``time-ordering'' mean in an acausal theory?
And might the wave packet of a particle traversing a CTC interfere with itself upon its simultaneous
emission and arrival?
We note that each of these two questions has been discussed before, the first one long ago in~\cite{Feynman-Wheeler},
and the second one more recently in~\cite{Greenberger}.
We offer no new insights into these questions.
Rather, we have been careful to paint a mainly classical picture in this paper.
We are content for now to let experiment be the arbiter of whether acausality is realizable in Nature.


Finally, we would like to conclude with some speculations.
In special relativity, space and time are unified.
However, it seems that there is still an implicit difference between space and time. The reason is that traveling
backwards in space appears to be easy, while traveling backwards in time requires a superluminal velocity.
So the question arises: why is there an apparently inexorable arrow of time in our universe?
The issue of chronology protection may somehow be related to the very concept of time.
Further theoretical investigations are badly needed.

While string theory \cite{PolchinskiBook} and loop quantum gravity \cite{Rovelli} are formulated very differently, there is a common
vision among them. Namely, a true theory of quantum gravity should be somehow background independent. This implies that spacetime is
actually a derived concept and hence emergent \cite{Seiberg}. In particular, the AdS/CFT correspondence \cite{Maldacena} suggests that
gravity is emergent. As observed in \cite{Polchinski}, the crucial point is that diffeomorphism invariance simply characterizes the
redundancies in the description of the gravity theory. But spacetime coordinates are themselves part of the redundant description in
general relativity. Thus, the emergence of a unique gravity requires the emergence of spacetime as well.
If the true quantum theory of gravity is indeed background independent and hence spacetime is emergent, then
the idea of CTCs or time travel is completely meaningless at the energy scale of quantum gravity, since
there is no spacetime at all. In this case, one can loosely say that chronology is ``trivially protected"
in that time is simply undefined.
The discussion of chronology protection and time travel then become intimately related with
the dynamics of how spacetime emerged.

If it turns out that the fundamental Planck scale is around a TeV as proposed by ADD,
then the LHC would be at the right
energy scale to elucidate our understanding of extra dimensions.
If it further turns out that Nature chooses an extra-dimensional metric which admits CTCs,
then discovery of acausal correlations at the LHC
would offer a fantastic new insight into the nature of spacetime.


\begin{acknowledgments}
We sincerely thank Alan Guth, Djordje Minic, Sandip Pakvasa, Tom Rizzo, Arkady Vainshtein,
and especially Andrew Tolley and the anonymous referee for useful comments.
This work was supported by US DOE grant DE-FG05-85ER40226.
\end{acknowledgments}

\appendix

\section{~~Covariant Approach to Light-Cone Analysis}
\label{app:GuthTime}
If a global time coordinate can be defined for a metric, then the metric cannot
contain CTC solutions.
A time-function $t$ is a global time coordinate if its four-gradient $\partial_\mu t$
is everywhere time-like, i.e., if $|\partial_\mu t|^2>0$~everywhere.
This covariant condition for the absence or presence of a CTC has been brought to our attention
by A. Guth.  He references a proof of this theorem in Ch.\ 8 of Wald's textbook~\cite{Wald1984gr}.
Here we wish to show that for our simple metric, this condition reduces to the light-cone
condition of Eq.~\rf{slopes1}.

The condition for the absence of any CTC is that
\beq{gtime1}
 |\partial_\mu t|^2=\partial_\mu t\,\partial_\nu t\,g^{\mu\nu} = s_\mu s_\nu  g^{\mu\nu} > 0 {\ \rm everywhere}\,,
 \quad{\rm with\ }s_\mu\equiv \partial_\mu t\,.
\eeq
Each of the four $s_\mu$ is the slope of the particle's world-line in the $\mu$-direction.
(Note that $s_\mu$ is {\sl not} a covariant four-vector.)
In this paper, we have chosen the time function to be the coordinate time $t$.
In addition, we have time-translation invariance along the brane directions,
but an off-diagonal metric element $g_{tu}$ in the time-bulk plane.
This off-diagonal element mixes $t$ and $u$,
leading to a nontrivial world-line $t(u)$ (see Eq.~\rf{tu}).
Thus, for our metric, Eq.~\rf{gtime1} becomes
\beq{gtime2}
|\partial_\mu t|^2 = g^{tt}+2s_u\,g^{tu} - s^2_u\,g^{uu}\,.
\eeq
Recalling that $g^{\mu\nu}$ is the matrix inverse of the metric $g_{\mu\nu}$,
one readily finds for the $2\times 2$ $t$-$u$~subspace that
$g^{tt}=-g_{uu}=+h(u)$, $g^{uu}=-g_{tt}=-1$, and $g^{tu}=+g_{tu}=g(u)$.
Substitution of these elements into Eq.~\rf{gtime2} then gives
\beq{gtime3}
|\partial_\mu t|^2 = -s_u^2 + 2g(u)\,s_u +h(u)\,.
\eeq
CTCs are allowed iff $|\partial_\mu t|^2$ passes through zero.
The quadratic form in Eq.~\rf{gtime3} may be written in terms of its roots $s_\pm$ as
$-(s_u-s_-)(s_u-s_+)$.  Comparing the two quadratic forms then gives
\beq{slopes3}
s_+ + s_- = 2g_{tu}(u) \quad{\rm and}\quad s_+ s_- = -h(u)\,.
\eeq
Thus, we are led via the covariant pathway to the massless-particle analog of
our intuitive light-cone slope condition for CTCs, given in Eq.~\rf{slopes2}.
(The sign of $g(u)$ is inconsequential since it can be reversed by simply redefining $u\rarr -u$.)

\section{~~ Divergence--Induced Operators and the Effective Cutoff}
\label{app:efftheory}
In $D$ dimensions, the $N$-pair Higgs operator
is proportional to
\beq{NonRen1}
(H^\dag H)^N_{\rm operator}\propto
\prod^N \lambda_4 \int d^D x \;\delta^{(D-4)}(\vx_\perp)\;\Delta_F (x-y)\,,
\eeq
with $\vx_\perp$ being the coordinates orthogonal to the brane.
The delta function in the integrand puts the operator on the brane where the $H$-field is nonzero.
The inter-connected spacetime propagators $D_F$ are
\beq{Delta}
\Delta_F (x-y) = \int d^D k \frac{e^{ik\cdot (x-y)}}{(k^2-m^2)}\,.
\eeq
Spacetime integrations lead to $N$ 4-dimensional delta-functions,
each enforcing  four-momentum conservation at one of the $N$ vertices.
Finally, these 4D~delta-functions may be integrated away to leave a single 4D~delta-function enforcing overall
momentum conservation, times $\lambda_4^N$ times the following schematic product of integrals:
\beq{NonRen2}
\int \frac{d^Dk}{(k^2-m^2)} \prod_{j=1}^{N-1} \int \frac{d^{(D-4)}{\vk_{j\perp}}}{(k_j^2-m^2+\cdots)}\,,
\eeq
where $\vk_\perp$ are the $\phi$-field momentum components orthogonal to the brane,
and the first four (``brane'') components of the $k_j$'s are fixed by the delta functions.
This integral product diverges as the $[4+N(D-6)]^{th}$ power of the cutoff $\Lambda$.
For example, in 5~dimensions, the divergence is quadratic for $N=2$ and logarithmic for $N=3$
(odd powers of divergence are removed by the symmetric integration that follows a Wick rotation).
In general, with more Higgs pairs or with more space-dimensions, the divergence is worse.
Consequently, the model is an effective theory, valid up to an energy cutoff of characteristic scale
$\Lambda\sim 1/(\lambda_4 \,L)$.

\section{~~Momentum in the Bulk Direction}
\label{app:p5}
In this Appendix, we wish to discuss the occurrence of conserved energy and brane three-momentum
for the particle, and the non-conserved particle momentum in the bulk direction, $u$.
Although the bulk momentum $p_5$ (and associated $p^5=G^{5\beta}\,p_\beta$) is neither conserved
nor observable, it is mathematically interesting in its own right.

The geodesic equation may be written as
\beq{geodsicEq}
\dot{\xi}_A = \half (\partial_A\,G_{BC})\,\xi^B\,\xi^C\,,
   \quad {\rm where\ } \xi^A\equiv \dot{x}^A
\eeq
is the tangent vector.
In this form, the geodesic equation makes it clear that for each Killing vector $\partial_A$
(i.e., $\partial_A$ such that $\partial_A \,G_{BC}=0,~ \forall\ \mbox{\footnotesize\sl{B,C}}$),
there is a conserved quantity $\xi_A=\dot{x}_A$.
Note that the conserved quantity carries a covariant (lower) index,
rather than a contravariant (upper) index.
For the 4D Minkowski metric, this is a moot point since upper and lower indices are simply related
by $\pm 1$.  However, for a more general metric, this point is crucial.

Since our metric depends only on $u$, it admits four conserved quantities.
These are $\dot{x}_0 = G_{0A}\,\dot{x}^A = \dot{t}+g(u)\,\dot{u}$, and
$\dot{x}_j= G_{jA}\,\dot{x}^A = -\dot{x}^j=-p^j/m$, i.e.,
the three-momentum $\vec{p}$ on the brane is conserved.
The conserved quantity $\dot{t}+g(u)\,\dot{u}$ resulting from the Killing vector $\partial_0$
must be proportional to the eigenvalue of the generator of time translation, i.e., the energy operator.
We derived the energy eigenvalue $E_n$ in the main text, and now we equate the two.
Using the initial value for conserved $\dot{t}+g(u)\,\dot{u}$,
we have
\beq{quantized}
E_n = m\,(\gamma_0+g_0\,\dot{u}_0)\,.
\eeq
That the mass $m$ is the proportionality constant is readily determined by taking the
4D limit of this equation, i.e., setting~$n$ and $\bar{g}$ to zero in $E_n$ on the LHS, and $\dot{u}_0$ to zero on the RHS.
In this paper, we do not exploit the relation~\rf{quantized}.

The momentum in the bulk direction is not conserved, owing to the breaking of the translational
invariance in the $u$~direction by the brane.
Nevertheless, we may use the relation between momentum operator and
generator for infinitesimal space translations to define it.
The momentum then satisfies the standard Dirac commutator with its conjugate variable, $x_5$.
The momentum operator expressed in position space then becomes $P_5 = -i\,\partial_5$.
Operating on the $n^{th}$-mode KG plane wave then determines its eigenvalue to be
\beq{p_5}
p_5 = -E_n\,(g(u)-\bar{g}) +\frac{n}{R}\,.
\eeq
It is obvious that $p_5$ is non-conserved, because $g(u)$ varies with $u$
whereas all other terms in $p_5$ are constants or conserved quantities.
The fact that $p_5$ depends on the global element $\bar{g}$ is an expression of ``awareness''
of the periodic boundary condition in the $u$~direction.
It then follows that $E_n$ is also aware of the boundary condition through $\bar{g}$, because
$E_n$ depends on $p_5$ (as well as on $\vec{p}$).

We conclude this Appendix by noting that the value of $p_5$ averaged over a cycle in $u$ is
\beq{ave_p_5}
\overline{p_5}\equiv \frac{1}{L}\int_0^L du\;p_5 = \frac{n}{R}\,,
\eeq
as  might have been expected by one familiar with compactified extra dimensions having a diagonal (possibly warped) metric.
An implication of this result is that any observable on the brane will depend on $p_5$ only
through a $\frac{n}{R}$ term.  Examples are the energy eigenvalues $E_n$, which are invariants
and so have the same value on or off the brane.
Thus, they may depend on $p_5$ only through $\frac{n}{R}$, and they do.

\section{~~ SM Higgs Decay:~ $h\rightarrow \bar{\phi}_{n_{1}} \,\bar{\phi}_{n_{2}}$}
\label{app:SMHiggsDecay}
In this Appendix, we calculate the decay width of a SM Higgs doublet $h$ of mass $M_h$ to a pair
of singlets $\phi_{n_1}$ and $\phi_{n_2}$.
Here it is assumed that the states $h$ and $\phi_n$ are not mixed.
Such is the case if a $\phi_n\leftrightarrow -\phi_n$ $Z_2$--symmetry is imposed.

Care is needed to correctly incorporate the unusual energy dispersion formula $E_n$ of Eq.~\rf{E_n},
and the compactified nature of the extra dimension.
The Lorentz invariant integral $\int\, \frac{d^3 \vec{p}}{2 \,E}=\int d^4 p\;\delta(p^2-m^2)\,\Theta(p^0)$,
appropriate for flat Minkowski space must be promoted to a covariant integral.
In principle, $d^4p$ is made covariant by multiplying it with $\sqrt{|{\rm Det(G_{AB})}|}$.
However, decay of the SM Higgs occurs only on the brane, so it is the determinant of the induced 4D metric
that enters here, and the induced 4D metric is nothing but the familiar Minkowski metric with ${|\rm Det}|=1$.
Thus $d^4p$ is invariant.  In addition, the quadratic form in the delta function is the eigenvalue of the
scalar Klein-Gordon operator $G^{AB}\partial_A\,\partial_B +m^2$, so it too is invariant.
Thus, the correct, invariant phase space integral is
\bea
\label{correctPS}
&& \sum_n\,\int\,d^4 p \; \delta\left(\,(1-\bar{g}^2)\,p_0^2-2\,\bar{g}\,p_0\,\frac{n}{R}
          -\vec{p}^{\ 2}-\frac{n^2}{R^2}-m^2\,\right)\; \theta(p_0) \nonumber \\
&&=
\sum_n\,\int\, \frac{d^3 \vec{p}}{2 \,[\,(1-\bar{g}^2)\,E_n-\frac{\bar{g}\,n}{R}\,]}
  = \sum_n\,\int\,\frac{d^3\vec{p}}{2\,\sqrt{(1-\bar{g}^2)(\vec{p}^{\ 2}+m^2)+\left(\frac{n}{R}\right)^2}}\,.
\eea
Here the argument of the delta function is just the quadratic form of the dispersion relation
(the generalization of the 4D Minkowski space dispersion relation $E^2=\vec{p}^{\ 2} +m^2$)
given in Eq.~\rf{qdispersion}, and the latter equality follows from Eq.~\rf{E_n}.
The periodic boundary condition in the $u$~direction
enters the dispersion relation through the mean metric element $\bar{g}$.

Our calculation below for the decay width $\Gamma_{h \rightarrow \bar{\phi}_{n_{1}}\, \bar{\phi}_{n_2}}$
follows the treatment given in Section 4.5 of \cite{Peskin}.
Translational invariance in time (for any time-independent metric) guarantees energy conservation,
and translational invariance in space along the three brane directions
guarantees three-momentum conservation.
Thus, the tree-level decay width in the center-of-momentum frame is given by
\bea
\label{App:width}
&&\Gamma_{h \rightarrow \bar{\phi}_{n_{1}}\, \bar{\phi}_{n_{2}}} \nonumber \\
&& \nonumber \\
&&=\frac{1}{2\,E_{\textrm{cm}}}\,
\int\, \frac{d^3 \vec{p}_1}{(2\pi)^3\,2 \,[\,(1-\bar{g}^2)\,E_{n_1}-\frac{\bar{g}\,n_1}{R}\,]} \,
\int\, \frac{d^3 \vec{p}_2}{(2\pi)^3\,2 \,[\,(1-\bar{g}^2)\,E_{n_2}-\frac{\bar{g}\,n_2}{R}\,]} \nonumber \\
&&~~~~~~~~~~~~~~~~~~~~~~~~~~~~~~~~~~~ \times|\mathcal{M}(h \rightarrow \bar{\phi}_{n_{1}}\, \bar{\phi}_{n_{2}})|^2\;
     (2\pi)^4\; \delta^{(3)}(\,\vec{p}_1+\vec{p}_2\,) \,\delta(\,E_{\textrm{cm}}-E_{n_1}-E_{n_2}\,) \nonumber \\
&& \nonumber \\
&&=\frac{\la_4^2\,v^2}{2\,E_{\textrm{cm}}}\,\int\, \frac{d|\vec{p}_1|\; |\vec{p}_1|^2\; d \Omega}{(2\pi)^3\;
     2\,[\,(1-\bar{g}^2)\,E_{n_1}-\frac{\bar{g}\,n_1}{R}\,]\,2\,[\,(1-\bar{g}^2)\,E_{n_2}-\frac{\bar{g}\,n_2}{R}\,] }
\,(2\pi)\,\delta(\,E_{\textrm{cm}}-E_{n_1}-E_{n_2}\,) \nonumber \\
&& \nonumber \\
&& = \frac{\la_4^2\,v^2}{2\,E_{\textrm{cm}}}\,\int\,d\Omega\;
     \frac{|\vec{p}|^2}{16\,\pi^2\,[\,(1-\bar{g}^2)\,E_{n_1}-\frac{\bar{g}\,n_1}{R}\,]\,[\,(1-\bar{g}^2)\,E_{n_2}-\frac{\bar{g}\,n_2}{R}\,] }\,
     \nonumber \\
     &&~~~~~~~~~~~~~~~~~~~~~~~~~~~~~~~~~~~ \left(\,\frac{|\vec{p}|}{(1-\bar{g}^2)\,E_{n_1}-\frac{\bar{g}\,n_1}{R}}+\frac{|\vec{p}|}
     {(1-\bar{g}^2)\,E_{n_2}-\frac{\bar{g}\,n_2}{R}}\,\right)^{-1}\nonumber\\
&& \nonumber \\
&& = \frac{\la_4^2\,v^2}{8\pi\,E_{\textrm{cm}}}\,
     \left( \frac{|\vec{p}|}{(1-\bar{g}^2)\,E_{n_1}-\frac{\bar{g}\,n_1}{R} + (1-\bar{g}^2)\,E_{n_2}-\frac{\bar{g}\,n_2}{R} } \right)\,.
\eea
Using the condition $E_{\textrm{cm}}= M_h = E_{n_1}(|\vec{p}|) +E_{n_2}(|\vec{p}|)$
and our dispersion relation, we arrive at
\bea
\label{GeneralWidth}
\Gamma_{h \rightarrow \bar{\phi}_{n_{1}}\, \bar{\phi}_{n_{2}}} = \frac{\la_4^2\,v^2}{16 \,\pi\,M_h }\;\beta_{n_1,n_2}\,,
\eea
where
\bea
\label{trianglefcn1}
&&\beta_{n_1,n_2} \nonumber \\ &=&
\frac{\sqrt{\left(\,(1-\bar{g}^2)\,M_h-\frac{\bar{g}\,(n_1+n_2)}{R}\,\right)^4-2\left(\,(1-\bar{g}^2)\,M_h-\frac{\bar{g}\,(n_1+n_2)}{R}\right)^2\;
\left(\,\bar{M}_{n_{1}}^2+\bar{M}_{n_{2}}^2\,\right)+\left(\,\bar{M}_{n_{1}}^2-\bar{M}_{n_{2}}^2\,\right)^2 }}
{\sqrt{1-\bar{g}^2}\,\left(\,(1-\bar{g}^2)\,M_h-\frac{\bar{g}\,(n_1+n_2)}{R}\,\right)^2} \nonumber \\
\eea
and $\bar{M}_{n_j}^2 \equiv (1-\bar{g}^2)\,m^2+(\frac{n_j^2}{R^2})$.
Perhaps a more familiar form for $\beta_{n_1,n_2}$, obtained by rearrangement of terms,\footnote
{
The equivalence of the argument of the square root in Eq.~\rf{trianglefcn1}
to the triangle function is easily seen by noting that the former is of the form $A^2-2A(B+C)+(B-C)^2$,
which when expanded explicitly displays the
symmetric form of the triangle function, $\lambda (A,B,C) = A^2 +B^2 +C^2 -2AB-2BC-2AC$.
A further feature of the triangle function, useful for extracting $\vec{p}^{\ 2} (E)$, is that
$\lambda (A,B+\vec{p}^{\ 2} ,C+\vec{p}^{\ 2} )=0$ implies that $\vec{p}^{\ 2} = \frac{\lambda (A,B,C)}{4A}$.
}
is
\beq{trianglefcn2}
\beta_{n_1,n_2}
= \frac{ \sqrt{\large{\lambda}\left(\,[\,(1-\bar{g}^2)\,M_h-\frac{\bar{g}\,(n_1+n_2)}{R}\,]^2, \bar{M}_{n_{1}}^2, \bar{M}_{n_{2}}^2\,\right)}}
     {\sqrt{1-\bar{g}^2}\,\left(\,(1-\bar{g}^2)\,M_h-\frac{\bar{g}\,(n_1+n_2)}{R}\,\right)^2}\,,
\eeq
where $\lambda(s,m_a^2,m_b^2)= (s-m_a^2-m_b^2)^2 -4 m_a^2 m_b^2$ is the usual triangle function employed
in flat space calculations.

Notice that when the final-state particles are identical, there are two possible contractions in the amplitude rather than one,
and a reduction of the two-body phase space from a sphere to a hemisphere to avoid double counting of identical particles.
The net result is an extra factor of $2^2\times \half = 2$.

There is a subtlety associated with apparent momentum non-conservation in the $u$-direction.
The existence of the brane at $u=0$ breaks translational invariance in the $u$-direction, and so we should not
expect particle momentum in the $u$-direction to be conserved.
For $q\equiv (n_1+n_2)$, the particle momentum leaving the brane in the $u$-direction is $q/R$, $q\equiv (n_1+n_2)$.
The ``missing momentum'' $-q/R$ is absorbed by the recoil of the brane.
This is analogous to the
apparent lack of conservation of $z$-momentum when a child jumps upward from a surface,
either rigid like the Earth's surface or elastic like a trampoline's surface.
Presumably, a form factor $\Fb{q}$ which characterizes the dynamic response of the brane
is included in $|M|^2$ above, and arrives as a factor in Eq.~\rf{GeneralWidth}.
Only for $q=0$, i.e. for $n_2=-n_1$ does the brane not enter the dynamics, so $\Fb{0}=1$.
In this work, we adopt the rigid picture of the brane, in which the net momentum of the exiting Higgs singlet pair,
$q/R$, is so small compared to brane tension that $\Fb{q}\approx \Fb{0} = 1$ for all $q$.

Finally, we state the obvious, that the total width of the SM Higgs to singlet Higgs pairs is
\beq{total-width}
\Gamma_{h\rarr\phi\phi}=\sum_{n_1,\,n_2} \Gamma_{h\rarr \bar{\phi}_{n_1}\,\bar{\phi}_{n_2}}\,,
\eeq
where $\sum_{n_1,n_2}$ includes all pairs of modes which are kinematically allowed,
i.e.\ all pairs of mode numbers for which the $\lambda$-function in Eq.~\rf{trianglefcn2} is positive.


 \section{~~Higgs Singlet--Doublet Mixing}
 \label{app:mixing}

 From Eq.~\rf{L4D2}, the contribution of $\int du\,\mathcal{L}_I$ to the
 mass-squared matrix mixing the Higgs doublet and tower of singlet states is
 \beq{massmatrix}
 {\cal M}^2 =
 \left(
 \ba{ccccc}
 M_h^2  & \la_3v         & \la_3v   & \la_3v   & \cdots \\
 \la_3v & M_0^2+\la_4v^2 & \la_4v^2 & \la_4v^2 & \cdots \\
 \la_3v & \la_4v^2 & M_1^2+\la_4v^2 & \la_4v^2 & \cdots \\
 \vdots & \vdots   & \la_4v^2       & \ddots   & \la_4v^2 \\
 \ea
 \right)\,.
 \eeq
(We do not consider here the mixing contribution from $\int du\,\mathcal{L}_0$.)
 Subtracting $\lambda \mathds{1} $ from this matrix and taking the determinant then gives the secular equation for the
 mass-squared eigenvalues $\lambda$.

 We may use Schur's determinant equation to simplify the calculation. For a matrix of the form
 \bea
 \mathds{M}=\left(
     \begin{array}{cc}
       \mathds{A}_{p\times p} & \mathds{B}_{p\times q} \\
       \mathds{C}_{q\times p} & \mathds{D}_{q\times q} \\
     \end{array}
   \right)\,,
 \eea
 the determinant of $\mathds{M}$ is given by $\rm{Det} (\mathds{M})=\rm{Det}\,(\mathds{D}) \,
 \rm{Det}(\mathds{A}-\mathds{B}\, \mathds{D}^{-1}\, \mathds{C})$.
 We choose $\mathds{A}$ to be the first entry in the upper left corner, and work
 in 0$^{\rm{th}}$ order of $\la_4\,v^2$.
 Schur's form then implies that
 \bea
 0=\textrm{Det} ({\cal M}^2- \lambda \mathds{1}) = \left[\,\prod_{n=0}^{\infty}\, (M_n^2-\la)\,\right]\,\left[\,(M_h^2-\la)\,
 -(\la_3\,v)^2\,\sum_{n=0}^{\infty}\,\frac{1}{(M_n^2-\la)}\,\right]\,.
 \eea

 If we are interested in the mixing of the Higgs doublet with a particular singlet mode $\bar{\phi}_n$,
 we may organize the secular equation as
 \beq{secular1}
 0= \left[\,(M_h^2-\la)\,(M_n^2-\la)-(\la_3\,v)^2\,\right]
    - (\la_3\,v)^2\left[\,\sum_{q\ne n}^{\infty}\,\frac{(M_n^2-\la)}{(M_q^2-\la)}\,\right]\,.
 \eeq
 For small enough values of $\la_3$, one may argue that the mass-squared eigenvalue $\la_n$ for the perturbed state $\bar{\phi}_n$
 remains sufficiently close to $M_n^2$ that $\left[\,\sum_{q\ne n}^{\infty}\,\frac{(M_n^2-\la)}{(M_q^2-\la)}\,\right]\,$
 may be neglected.
 For this case, the mixing angle between states $h$ and $\bar{\phi}_n$ becomes $\tan 2\thet\sim \frac{2\,\la_3\,v}{|M_n^2-M_h^2|}$.
 For larger values of $\la_3$, or large values of $\la_4$, more care would be needed.

 To quantify these remarks, we first solve the piece of the secular equation in the first bracket
 of~\rf{secular1} to get
\beq{evals}
 \la_n=\half\left[ M_n^2+M_h^2 +\sqrt{ (M_n^2-M_h^2)^2 +(2\,\la_3\,v)^2}\right]
 \stackrel{\la_3\,v\,\ll\,(M_n^2-M_h^2)}{\longrightarrow} M_n^2+\frac{(\la_3\,v)^2}{(M_n^2-M_h^2)}\,.
 \eeq
 Then, we insert this perturbative result back into Eq.~\rf{secular1} to evaluate the residual given by the second bracket.
 The result is
 \beq{residual}
 {\rm residual}=(\la_3\,v)^4\,\sum_{q\neq n}
    \left[\,\left(\frac{q^2}{R^2}-\frac{n^2}{R^2}\right)\,(M_n^2-M_h^2)-(\la_3\,v)^2\,\right]^{-1}\,.
 \eeq

 Thus, for $\la_3\,v\ll(M_n^2-M_h^2)$, the residual is a negligible order $(\la_3\,v)^4$ perturbation,
 and the results $\la_n \approx M_n^2+\frac{(\la_3v)^2}{(M_n^2-M_h^2)}$ and
 $\tan 2\thet\sim \frac{2\,\la_3\,v}{|M_n^2-M_h^2|}$ are robust.



\begin{thebibliography}{99}

\bibitem{vanStockum} W. J. van Stockum, ``\emph{Gravitational field of a distribution of particles rotating around an
axis of symmetry}," Proc. R. Soc. Edin. \textbf{57}, 135 (1937).

\bibitem{Tipler} F. J. Tipler, ``\emph{Rotating cylinders and the possibility of global causality violation}," Phys.
Rev. D \textbf{9}, 2203-2206 (1974).

\bibitem{Godel} K. G\"odel, ``\emph{An Example Of A New Type Of Cosmological Solutions Of Einstein's Field
Equations Of Gravitation}," Rev. Mod. Phys. \textbf{21}, 447 (1949).

\bibitem{Wheeler} J. A. Wheeler, ``\emph{ Geons}," Phys. Rev. \textbf{97}, 511 (1955); J. A. Wheeler, ``\emph{On the nature of
quantum geometrodynamics}," Annals Phys. \textbf{2}, 604 (1957).

\bibitem{Kerr} S. W. Hawking, G. F. R. Ellis, ``\emph{The Large Scale Structure of Spacetime}," Cambridge
University Press, New York, 1973.

\bibitem{MTY} M. S. Morris and K. S. Thorne, ``\emph{Wormholes In Spacetime And Their Use For Interstellar
Travel: A Tool For Teaching General Relativity}," Am. J. Phys. \textbf{56}, 395 (1988);
M. S. Morris, K. S. Thorne and U. Yurtsever, ``\emph{Wormholes, Time Machines, And The
Weak Energy Condition}," Phys. Rev. Lett. \textbf{61}, 1446 (1988).

\bibitem{Gott} J. R. I. Gott, ``\emph{Closed Timelike Curves Produced By Pairs Of Moving Cosmic Strings:
Exact Solutions}," Phys. Rev. Lett. \textbf{66}, 1126 (1991).

\bibitem{warp} M. Alcubierre, ``\emph{The warp drive: hyper-fast travel within general relativity}," Class. Quant.
Grav. \textbf{11}, L73 (1994) [arXiv:gr-qc/0009013]; A. E. Everett, ``\emph{Warp drive and causality},"
Phys. Rev. D \textbf{53}, 7365 (1996).

\bibitem{Ori} A. Ori, ``\emph{A new time-machine model with compact vacuum core}," Phys. Rev. Lett. \textbf{95}, 021101 (2005)
[arXiv:gr-qc/0503077].

\bibitem{Gron}
  O.~Gron and S.~Johannesen,
  ``\emph{A spacetime with closed timelike geodesics everywhere}," Nuovo Cim. B \textbf{125},
     1215 (2011) [arXiv:1004.3235 [gr-qc]];
  ibid.,
  ``\emph{Closed timelike geodesics in a gas of cosmic strings},"
  New J.\ Phys.\  {\bf 10}, 103025 (2008)
  [arXiv:gr-qc/0703139].

\bibitem{Visser} M. Visser, ``\emph{Lorentzian Wormholes - from Einstein to Hawking}," Springer, New York,
1996.

\bibitem{Hawking} S. W. Hawking, ``\emph{The Chronology protection conjecture}," Phys. Rev. D \textbf{46}, 603 (1992).

\bibitem{Wald} B.~S.~Kay, M.~J.~Radzikowski and R.~M.~Wald, ``\emph{Quantum field theory on spacetimes with a
compactly generated Cauchy horizon}," Commun.\ Math.\ Phys.\  {\bf 183}, 533 (1997) [arXiv:gr-qc/9603012].

\bibitem{Horava} E.~K.~Boyda, S.~Ganguli, P.~Horava and U.~Varadarajan, ``\emph{Holographic protection of chronology in
universes of the Goedel type}," Phys.\ Rev.\  D {\bf 67}, 106003 (2003) [arXiv:hep-th/0212087].

\bibitem{Dyson} L. Dyson, ``\emph{Chronology protection in string theory}," JHEP \textbf{0403}, 024 (2004)
[arXiv:hep-th/0302052].

\bibitem{AdS} M.~M.~Caldarelli, D.~Klemm and P.~J.~Silva, ``\emph{Chronology protection in anti-de Sitter},"
Class.\ Quant.\ Grav.\  {\bf 22}, 3461 (2005) [arXiv:hep-th/0411203].

\bibitem{Johnson} C.~V.~Johnson and H.~G.~Svendsen, ``\emph{An exact string theory model of closed time-like curves and
cosmological singularities}," Phys.\ Rev.\  D {\bf 70}, 126011 (2004) [arXiv:hep-th/0405141].

\bibitem{ADD}
N. Arkani-Hamed, S. Dimopoulos and G. R. Dvali, ``\emph{The hierarchy problem and new
dimensions at a millimeter}," Phys. Lett. B \textbf{429}, 263 (1998) [arXiv:hep-ph/9803315];
I.~Antoniadis, N.~Arkani-Hamed, S.~Dimopoulos and G.~R.~Dvali,
``\emph{New dimensions at a millimeter to a Fermi and superstrings at a TeV},"
Phys.\ Lett.\  B {\bf 436}, 257 (1998) [arXiv:hep-ph/9804398];
N. Arkani-Hamed, S. Dimopoulos and G. R. Dvali, ``\emph{Phenomenology,
astrophysics and cosmology of theories with sub-millimeter dimensions and
TeV scale quantum gravity}," Phys. Rev. D \textbf{59}, 086004 (1999) [arXiv:hep-ph/9807344].

\bibitem{shortcut1} G. Kaelbermann, ``\emph{Communication through an extra dimension}," Int. J. Mod. Phys. A \textbf{15},
3197 (2000) [arXiv:gr-qc/9910063].

\bibitem{shortcut2} H. Ishihara, ``\emph{Causality of the brane universe}," Phys. Rev. Lett. \textbf{86}, 381 (2001)
[arXiv:gr-qc/0007070].

\bibitem{shortcut3} R. R. Caldwell and D. Langlois, ``\emph{Shortcuts in the fifth dimension}," Phys. Lett. B \textbf{511},
129 (2001) [arXiv:gr-qc/0103070].

\bibitem{shortcut4} H. Stoica, ``\emph{Comment on 4D Lorentz invariance violations in the brane-world}," JHEP
\textbf{0207}, 060 (2002) [arXiv:hep-th/0112020].

\bibitem{shortcut5} E. Abdalla, A. G. Casali and B. Cuadros-Melgar, ``\emph{Shortcuts in Cosmological Branes},"
Int. J. Theor. Phys. \textbf{43}, 801 (2004) [arXiv:hep-th/0501076].

\bibitem{RS} L. Randall and R. Sundrum, ``\emph{A large mass hierarchy from a small extra dimension},"
Phys. Rev. Lett. \textbf{83}, 3370 (1999) [arXiv:hep-ph/9905221]; L. Randall
and R. Sundrum, ``\emph{An alternative to compactification}," Phys. Rev. Lett. \textbf{83}, 4690
(1999) [arXiv:hep-th/9906064]

\bibitem{csaki} C. Csaki, J. Erlich and C. Grojean, ``\emph{Gravitational Lorentz violations and adjustment
of the cosmological constant in asymmetrically warped spacetimes}," Nucl. Phys. B \textbf{604},
312 (2001) [arXiv:hep-th/0012143].

\bibitem{freese} D. J. H. Chung and K. Freese, ``\emph{Cosmological challenges in theories with extra dimensions
and remarks on the horizon problem}," Phys. Rev. D \textbf{61}, 023511 (1999)
[arXiv:hep-ph/9906542]; D. J. H. Chung and K. Freese, ``\emph{Can geodesics in extra dimensions
solve the cosmological horizon problem?}," Phys. Rev. D \textbf{62}, 063513 (2000)
[arXiv:hep-ph/9910235].

\bibitem{Tom} H.~Pas, S.~Pakvasa, J.~Dent and T.~J.~Weiler, ``\emph{Closed timelike curves in asymmetrically warped brane universes},"
Phys.\ Rev.\  D {\bf 80}, 044008 (2009) [arXiv:gr-qc/0603045].

\bibitem{Novikov:1989sd} I.~D.~Novikov, ``\emph{Time machine and selfconsistent evolution in problems with selfinteraction},"
Phys.\ Rev.\  D {\bf 45}, 1989 (1992).

\bibitem{Carlini}
  A.~Carlini, V.~P.~Frolov, M.~B.~Mensky, I.~D.~Novikov and H.~H.~Soleng,
  ``\emph{Time machines: The Principle of selfconsistency as a consequence of the
  principle of minimal action},"
  Int.\ J.\ Mod.\ Phys.\  D {\bf 4}, 557 (1995)
  [Erratum-ibid.\  D {\bf 5}, 99 (1996)]
  [arXiv:gr-qc/9506087].

\bibitem{Echeverria:1991nk} F.~Echeverria, G.~Klinkhammer and K.~S.~Thorne,
``\emph{Billiard balls in wormhole space-times with closed timelike curves: Classical theory},"
Phys.\ Rev.\  D {\bf 44}, 1077 (1991).

\bibitem{Friedman:1990xc} J.~Friedman, M.~S.~Morris, I.~D.~Novikov, F.~Echeverria, G.~Klinkhammer, K.~S.~Thorne and U.~Yurtsever,
``\emph{Cauchy Problem In Space-Times With Closed Timelike Curves},"
Phys.\ Rev.\  D {\bf 42} (1990) 1915.

\bibitem{Friedman:1992jc} J.~L.~Friedman, N.~J.~Papastamatiou and J.~Z.~Simon,
``\emph{Failure of unitarity for interacting fields on space-times with closed
timelike curves}," Phys.\ Rev.\  D {\bf 46}, 4456 (1992).

\bibitem{Boulware:1992pm} D.~G.~Boulware,
``\emph{Quantum Field Theory In Spaces With Closed Timelike Curves}," Phys.\ Rev.\  D {\bf 46}, 4421 (1992)
[arXiv:hep-th/9207054].

\bibitem{Mironov:2007bm} A.~Mironov, A.~Morozov and T.~N.~Tomaras, ``\emph{If LHC is a Mini-Time-Machines Factory, Can We Notice?},"
Facta Univ.\ Ser.\ Phys.\ Chem.\ Tech.\  {\bf 4}, 381 (2006) [arXiv:0710.3395 [hep-th]].

\bibitem{Hartle:1993sg} J.~B.~Hartle, ``\emph{Unitarity and causality in generalized quantum mechanics for nonchronal space-times},"
Phys.\ Rev.\  D {\bf 49}, 6543 (1994) [arXiv:gr-qc/9309012].

\bibitem{Gielen}
  S.~Gielen,
  ``\emph{Comment on ``Causality-violating Higgs singlets at the LHC"},''
  Phys.\ Rev.\ D {\bf 88}, 068701 (2013)
  [arXiv:1302.1711 [hep-ph]].


\bibitem{DJtH-Annals}
 S.~Deser, R.~Jackiw and G.~'t~Hooft,
  ``\emph{Three-Dimensional Einstein Gravity: Dynamics Of Flat Space},"
  Annals Phys.\  {\bf 152}, 220 (1984).

\bibitem{DJ-Feinberg}
  S.~Deser and R.~Jackiw,
  ``\emph{Time travel?},"
  Comments Nucl.\ Part.\ Phys.\  {\bf 20}, 337 (1992)
  [arXiv:hep-th/9206094].

\bibitem{Hooft} S.~Deser, R.~Jackiw and G.~'t Hooft, ``\emph{Physical cosmic strings do not generate closed timelike curves},"
  Phys.\ Rev.\ Lett.\  {\bf 68}, 267 (1992).

\bibitem{Shore}
 G.~M.~Shore,
 ``\emph{Constructing time machines},'
  Int.\ J.\ Mod.\ Phys.\  A {\bf 18}, 4169 (2003)
  [arXiv:gr-qc/0210048], and references therein.

\bibitem{Carroll} S.~M.~Carroll, E.~Farhi, A.~H.~Guth and K.~D.~Olum,
``\emph{Energy momentum restrictions on the creation of Gott time machines},"
Phys.\ Rev.\  D {\bf 50}, 6190 (1994) [arXiv:gr-qc/9404065].

\bibitem{Tye} B.~Shlaer and S.~H.~Tye, ``\emph{Cosmic string lensing and closed time-like curves},"
Phys.\ Rev.\  D {\bf 72}, 043532 (2005) [arXiv:hep-th/0502242].

\bibitem{Cullen} S.~Cullen and M.~Perelstein, ``\emph{SN1987A constraints on large compact dimensions},"
Phys.\ Rev.\ Lett.\  {\bf 83}, 268 (1999)
[arXiv:hep-ph/9903422]; R.~Franceschini, G.~F.~Giudice, P.~P.~Giardino, P.~Lodone and A.~Strumia,
  ``\emph{LHC bounds on large extra dimensions}," JHEP {\bf 05} (2011) 092 [arXiv:1101.4919 [hep-ph]] (which provides less stringent constraints derived directly from new
  LHC data).

\bibitem{Kribs} G.~D.~Kribs, ``\emph{Phenomenology of extra dimensions},"
arXiv:hep-ph/0605325.

\bibitem{Giudice}
  G.~F.~Giudice, R.~Rattazzi and J.~D.~Wells,
  ``\emph{Graviscalars from higher dimensional metrics and curvature Higgs mixing},"
  Nucl.\ Phys.\ B {\bf 595}, 250 (2001)
  [hep-ph/0002178].

\bibitem{Gunion}
 D.~Dominici and J.~F.~Gunion,
  ``\emph{Invisible Higgs Decays from Higgs Graviscalar Mixing},"
  Phys.\ Rev.\ D {\bf 80}, 115006 (2009)
  [arXiv:0902.1512 [hep-ph]].

\bibitem{Burgess} C.~P.~Burgess, M.~Pospelov and T.~ter Veldhuis, ``\emph{The minimal model of nonbaryonic dark matter: A singlet scalar},"
Nucl.\ Phys.\  B {\bf 619}, 709 (2001) [arXiv:hep-ph/0011335].

\bibitem{Bento} M.~C.~Bento, O.~Bertolami, R.~Rosenfeld and L.~Teodoro, ``\emph{Self-interacting dark matter and invisibly decaying Higgs},"
Phys.\ Rev.\  D {\bf 62}, 041302 (2000) [arXiv:astro-ph/0003350].

\bibitem{CDMS2} Z.~Ahmed {\it et al.}  [The CDMS-II Collaboration], ``\emph{Results from the Final Exposure of the CDMS II Experiment},"
Science {\bf 327}, 1619 (2010) [arXiv:0912.3592 [astro-ph.CO]].

\bibitem{Asano} M.~Asano and R.~Kitano, ``\emph{Constraints on Scalar Phantoms}," Phys.\ Rev.\  D {\bf 81}, 054506 (2010)
[arXiv:1001.0486 [hep-ph]].

\bibitem{Farina} M.~Farina, D.~Pappadopulo and A.~Strumia, ``\emph{CDMS stands for Constrained Dark Matter Singlet},"
Phys.\ Lett.\  B {\bf 688}, 329 (2010) [arXiv:0912.5038 [hep-ph]].

\bibitem{Zee} V.~Silveira and A.~Zee, ``\emph{Scalar Phantoms}," Phys.\ Lett.\  B {\bf 161}, 136 (1985).





\bibitem{Nielsen:2007ak} H.~B.~Nielsen and M.~Ninomiya, ``\emph{Search for Future Influence from L.H.C},"
Int.\ J.\ Mod.\ Phys.\  A {\bf 23}, 919 (2008) [arXiv:0707.1919 [hep-ph]].

\bibitem{Aref'eva:2007vk} I.~Y.~Aref'eva and I.~V.~Volovich, ``\emph{Time Machine at the LHC},"
Int.\ J.\ Geom.\ Meth.\ Mod.\ Phys.\  {\bf 05}, 641 (2008) [arXiv:0710.2696 [hep-ph]].

\bibitem{LeeWick}
E.~Alvarez, L.~Da Rold, C.~Schat and A.~Szynkman,
  ``\emph{Vertex Displacements for Acausal Particles: Testing the Lee-Wick Standard
  Model at the LHC},"
  JHEP {\bf 0910}, 023 (2009)
  [arXiv:0908.2446 [hep-ph]];
B.~Grinstein, D.~O'Connell and M.~B.~Wise,
  ``\emph{The Lee-Wick standard model},"
  Phys.\ Rev.\  D {\bf 77}, 025012 (2008)
  [arXiv:0704.1845 [hep-ph]].

\bibitem{Feynman-Wheeler}
 J.~A.~Wheeler and R.~P.~Feynman,
  ``\emph{Interaction with the absorber as the mechanism of radiation},"
  Rev.\ Mod.\ Phys.\  {\bf 17}, 157 (1945); J.~A.~Wheeler and R.~P.~Feynman,
  ``\emph{Classical electrodynamics in terms of direct interparticle action},"
  Rev.\ Mod.\ Phys.\  {\bf 21}, 425 (1949).

\bibitem{Greenberger}
D. M. Greenberger and K. Svozil,
Chapter 4 of ``\emph{Quo Vadis Quantum Mechanics?}," ed.\, A. Elitzur, S. Dolev and N. Kolenda, Springer Verlag, Berlin (2005),
and arXiv:quant-ph/0506027.




\bibitem{PolchinskiBook} J.~Polchinski, ``\emph{String theory. Vol. 1: An introduction to the bosonic string},''
{\it  Cambridge, UK: University Press (1998) 402 p}; ``\emph{String theory. Vol. 2: Superstring theory and beyond},''
{\it  Cambridge, UK: University Press (1998) 531 p}.

\bibitem{Rovelli} C.~Rovelli, ``\emph{Quantum Gravity},'' {\it  Cambridge, UK: University Press (2004) 455 p}.

\bibitem{Seiberg} N.~Seiberg, ``\emph{Emergent spacetime},'' arXiv:hep-th/0601234.

\bibitem{Maldacena} J.~M.~Maldacena, ``\emph{The large N limit of superconformal field theories and supergravity},''
Adv.\ Theor.\ Math.\ Phys.\  {\bf 2}, 231 (1998), [arXiv:hep-th/9711200].

\bibitem{Polchinski} H.~Elvang and J.~Polchinski, ``\emph{The quantum Hall effect on $R^4$},'' arXiv:hep-th/0209104.


\bibitem{Wald1984gr}
R.~M.~Wald,
``\emph{General Relativity},''
Chicago, USA: University Press (1984).

\bibitem{Peskin}
  M.~E.~Peskin and D.~V.~Schroeder, ``\emph{An Introduction To Quantum Field Theory}'',
Westview Press, Boulder, Colorado, 1995.



\end{thebibliography}
\end{document}